\documentclass[journal]{IEEEtran}
\usepackage{cite}

\ifCLASSINFOpdf
\else
\fi
\usepackage{amsmath}
\usepackage{fixltx2e}
\usepackage[latin1]{inputenc}
\usepackage{graphicx}

\newcommand{\cov}{\mathrm{Cov}}
\newcommand{\esp}{\mathrm{E}}
\newcommand{\pr}{\mathrm{p}}
\newcommand{\ud}{\mathrm{d}}
\newcommand{\var}{\mathrm{Var}}

\newcommand{\gcite}{\cite}
\newcommand{\gcitet}{\cite}

\newcommand{\TF}[1]{T_{#1} ( t, f )}
\newcommand{\phase}[1]{\theta_{#1} ( t, f )}
\newcommand{\TFT}{time-frequency transform}

\newcommand{\ITC}[1]{\mathrm{ITC}_{#1_{1:N}} ( t, f )}
\newcommand{\avgAMP}[1]{\mathrm{avgAMP}_{#1_{1:N}} ( t, f )}
\newcommand{\POWavg}[1]{\mathrm{POWavg}_{#1_{1:N}} ( t, f )}

\newcommand{\approxim}{^{\dagger}}
\newcommand{\TFa}[1]{T_{#1}\approxim ( t, f )}
\newcommand{\ITCa}[1]{\mathrm{ITC}_{#1_{1:N}}\approxim ( t, f )}
\newcommand{\avgAMPa}[1]{\mathrm{avgAMP}_{#1_{1:N}}\approxim ( t, f )}

\newcommand{\POWavga}[1]{\mathrm{POWavg}_{#1_{1:N}}\approxim ( t, f )}

\newcommand{\supplmat}[1]{Supplementary Material~\##1}

\begin{document}
%
\title{Time-frequency analysis of event-related brain recordings: Connecting power of evoked potential and inter-trial coherence}
%
%
%

\author{Jonas Benhamou,
        Michel Le Van Quyen,
        and~Guillaume Marrelec
\thanks{All authors are with the Laboratoire d'imagerie biomédicale, LIB, Sorbonne Université, CNRS, Inserm, Paris, France. J. Benhamou is also with the {\'E}cole Nationale Supérieure de Techniques Avanc{\'e}es, ENSTA, Saclay, France.
}
\thanks{Manuscript received ???; revised ???. Copyright (c) 2021 IEEE. Personal use of this material is permitted. However, permission to use this material for any other purposes must be obtained from the IEEE by sending an email to pubs-permissions@ieee.org.}}

%
%

\markboth{}%
{Benhamou \MakeLowercase{\textit{et al.}}: Connecting the averaged evoked response power and inter-trial coherence}
%



\maketitle

\begin{abstract}
 \textit{Objective.} In neuroscience, time-frequency analysis has been used to get insight into brain rhythms from brain recordings. In event-related protocols, one applies it to investigate how the brain responds to a stimulation repeated over many trials. In this framework, three measures have been considered: the amplitude of the transform for each single trial averaged across trials, avgAMP; inter-trial phase coherence, ITC; and the power of the evoked potential transform, POWavg. These three measures are sensitive to different aspects of event-related responses, ITC and POWavg sharing a common sensitivity to phase resetting phenomena. \textit{Methods.} In the present manuscript, we further investigated the connection between ITC and POWavg using theoretical calculations, a simulation study and analysis of experimental data. \textit{Results.} We derived exact expressions for the relationship between POWavg and ITC in the particular case of the S-transform of an oscillatory signal. In the more general case, we showed that POWavg and ITC are connected through a relationship that roughly reads POWavg $\approx$ avgAMP\textsuperscript{2} $\times$ ITC\textsuperscript{2}. This result was confirmed on simulations. We finally compared the theoretical prediction with results from real data. \textit{Conclusion.} We showed that POWavg and ITC are related through an approximate, simple relationship that also involves avgAMP. \textit{Significance.} The presented relationship between POWavg, ITC, and avgAMP confirms previous empirical evidence and provides a novel perspective to investigate evoked brain rhythms. It may provide a significant refinement to the neuroscientific toolbox for studying evoked oscillations.
\end{abstract}


%
\IEEEpeerreviewmaketitle

\section{Introduction}
%
%
%
\IEEEPARstart{I}{n} signal processing, time-frequency (TF) analysis is a generic term that encompasses a variety of methods---such as the short-time Fourier transform \gcite{Grochenig-2001}, wavelets \gcite{Mallat-1999}, and the S-transform \gcite{Stockwell-1996}---for the analysis of signals whose frequency features vary with time \gcite{Cohen_L-1995, Flandrin-1999}. All methods have in common that they map a one-dimensional real or complex signal $x ( t )$ into a two-dimensional complex-valued function $\TF{x}$, called time-frequency transform. This transform provides, for each time $t$ and frequency $f$, a modulus $| \TF{x} |$ (also called amplitude) and an argument $\arg [ \TF{x} ]$ or $\phase{x}$ (also called phase). The terms amplitude and phase are meant to emphasize the expected relation between the amplitude and phase of an oscillatory signal and the modulus and argument of its \TFT \gcite[Chap.~4]{Mallat-1999}. Time-frequency analysis has been used in a wide variety of research fields, from geophysics to engineering to medicine \gcite{Addison-2002}.
\par
In neuroscience, neuronal assemblies are hypothesized to generate oscillating electromagnetic fields within specific frequency ranges, also called brain rhythms. There has been overwhelming evidence that these brain rhythms can be captured by in-vivo brain recording techniques, from noninvasive methods, such as electroencephalography (EEG) or magnetoencephalography (MEG), to invasive methods, such as intracranial EEG (iEEG) or microelectrode recordings (measuring local field potentials, LFPs) \gcite{Basar-1999, Singer-1999, Glass-2001, Buzsaki-2004, Laufs-2010, Fries-2015}. A common procedure to investigate brain rhythms in brain recordings is the so-called event-related protocol, where one records how the brain responds to a given stimulation over many repetitions, called trials. For $N$ trials, one then has a collection of $N$ signals $x_n ( t )$, $n = 1, \dots, N$. The brain activity measured in response to each stimulation is called an event-related response. The exact interpretation of event-related responses remains an open issue in neuroscience. It was suggested that they may result from a variable combination of two phenomena: stimulus-evoked neuronal activity and stimulus-induced phase resetting of ongoing neuronal dynamics \gcite{Penny-2002, Makeig-2002, Makeig-2004, David-2006}. In the former, the stimulus leads to the activation of specific neuronal populations, which contribute to new time-locked oscillations at each trial, translating into new power contributions to the ongoing brain activity. In the latter, the stimulus resets the phase of pre-existing oscillations so that a portion of the ongoing activity becomes transiently phase-locked in a particular frequency range. To analyze brain recordings from event-related protocols and be able to discriminate between evoked responses and phase resetting / induced responses, increasingly sophisticated methods have been used, including time-frequency analysis \gcite{Varela-2001, Delorme-2004, Le_Van_Quyen-2007}. So far, three time-frequency measures have been considered. A first measure is the average time-frequency amplitude, i.e., the amplitude $| \TF{x_n} |$ of the transform for each single trial $n$ averaged across trials, \gcite[Chap.~9]{Hari-2017}
\begin{equation} \label{eq:avgAMP:def}
 \mathrm{avgAMP} = \frac{ 1 } { N } \sum_{ n = 1 } ^ N \left| \TF{x_n} \right|.
 \end{equation}
A second measure is the so-called inter-trial phase coherence (ITC), which quantifies the phase consistency of brain responses over experimental trials \gcite{Makeig-2004}
\begin{equation} \label{eq:ITC:def}
 \mathrm{ITC} = \left| \frac{1}{N} \sum_{ n = 1 }^ N e ^ { i \phase{x_n} } \right|.
\end{equation}
A last measure applies the \TFT\ to the evoked potential, i.e., the average through trials of all signals, and extracts its power: If $\overline{x_n} ( t )$ is the the signal averaged over all $N$ trials, then \gcite{Tallon-Baudry-1996}
\begin{equation} \label{eq:POWavg:def}
 \mathrm{POWavg} = \left| \TF{\overline{x_n}} \right| ^ 2.
\end{equation}
Importantly, these three measures are sensitive to different aspects of event-related responses: evoked responses for avgAMP, and phase resetting for ITC and POWavg \gcite{Tallon-Baudry-1996, Makeig-2004}. To date, with these three measures, evidence for both evoked and induced oscillations has been reported in multiple cognitive protocols generating time-locked responses to specific sensory, cognitive or motor events \gcite{Tallon-Baudry-1999, Fuentemilla-2006, Hanslmayr-2007}.
\par
We are here interested in the observed common sensitivity of ITC and POWavg to induced responses as well as the overall similarity of their outputs \gcite{Valencia-2006}. More precisely, we investigate the theoretical underpinning of this empirical evidence. Intuitively, we would expect a low ITC to correspond to a low POWavg, as distinct components with phases that are not consistent across trials (i.e., with low ITC) should cancel out in the averaging process, leading to an average signal with low POWavg. In the present manuscript, we further investigate the connection between ITC and POWavg using theoretical calculations and a simulation study. We show that POWavg is approximately equal to the product of avgAMP squared by ITC squared, 
\begin{equation} \label{eq:rel:approx}
 \mathrm{POWavg} \approx \mathrm{avgAMP} ^ 2 \times \mathrm{ITC}^2,
\end{equation}
under certain conditions. More precisely, we first derive several approximations of the relationship between POWavg, ITC and avgAMP in the particular case of the S-transform of an oscillatory signal with increasing levels of complexity. We then investigate the more general case and derive both the exact form of this relationship---see \eqref{eq:rel:esp}---, as well as the assumptions leading to the simpler form of \eqref{eq:rel:approx}. We show that this result also holds in the different setting of synthetic data simulating high frequency oscillations. We finally investigate the validity of this result for experimental data.
\par
The outline of the manuscript is the following. In Section~\ref{s:td}, we derive several variants of \eqref{eq:rel:approx}. In Section~\ref{s:ss}, we perform a time-frequency analysis of synthetic data originating from a simulation study. The real data are analyzed in Section~\ref{s:ated}. Further issues are discussed in Section~\ref{s:disc}.

\section{Theoretical developments} \label{s:td}

In this section, we provide the theoretical connection between ITC and POWavg, thus specifying the form of \eqref{eq:rel:approx}. We start by stating some basic results (Section~\ref{ss:sbr}). We then expose the general framework for time-frequency analysis (Section~\ref{ss:tfa}) and give some results regarding avgAMP, ITC, and POWavg (Section~\ref{ss:tqoi}). We then provide an intuition for the existence of a relationship (Section~\ref{ss:ifc}) and derive the its exact form, both in the particular case of the S-transform of an oscillatory signal (Section~\ref{ss:ioom}), and in the general case (Section~\ref{ss:ge}).

\subsection{Some basic results} \label{ss:sbr}

We need the following results. 

\subsubsection{Complex numbers}

First, for any complex number $z$, absolute value $| z |$ and conjugate $z ^ *$ are related by $| z | ^ 2 = z z ^*$. We also have $z + z ^ * = 2 \Re ( z )$, where $\Re( z )$ is the real part of $z$.

\subsubsection{Random variables}

If $K$ random variables $X_1, \dots, X_K$ are independent, then the expectation of their product is equal to the product of their expectations \gcite[\S2.2.3]{Anderson_TW-1958}
\begin{equation} \label{eq:esp:prop}
 \esp \left( \prod_{ k = 1 } ^ K X_k \right) = \prod_{ k = 1 } ^ K \esp \left( X_k \right).
\end{equation}
For two complex-valued random variables $X$  and $Y$, we have
\begin{equation} \label{eq:def:cov}
 \cov ( X, Y ) = \esp ( X Y ^ * ) - \esp ( X ) \esp ( Y ) ^ *.
\end{equation}
In particular, this entails that
\begin{equation} \label{eq:def:var}
 \var ( X ) = \esp \left( | X | ^ 2 \right) - \left| \esp ( X ) \right| ^ 2.
\end{equation}
\par
A circular random variable is a random variable that, like an angle, is defined on the unit circle, i.e., whose value is only relevant modulo $2 \pi$. A key quantity for any circular random variable $\theta$ is its circular mean, defined as
\begin{equation} \label{eq:moycirc:def}
 \esp \left( e ^ { i \theta } \right) = \int \pr ( \theta ) \, e ^ { i \theta } \, \ud \theta.
\end{equation}
The argument of $\esp \left( e ^ { i \theta } \right)$ is the mean angle or mean direction, while the modulus of $\esp \left( e ^ { i \theta } \right)$ is the mean resultant length. It lays between 0 and 1 and is a measure of concentration of $\pr ( \theta )$.
\par
A circular variable $\theta$ is said to follow a von Mises distribution with mean direction $\theta_0$ and concentration parameter $\kappa$, denoted $\theta \sim \mathrm{VonMises} (\theta_0,\kappa)$, if its distribution is given by \gcite[\S3.5.4]{Mardia-2000}
$$\pr ( \theta ) = \frac{ 1 } { 2 \pi I_0 ( \kappa ) } e ^ { \kappa \cos ( \theta - \theta_0 ) },$$
where $I_0 ( \kappa )$ is the modified Bessel function of order 0,
\begin{equation}
 I_0 ( \kappa ) = \frac{ 1 } { 2 \pi } \int_{ - \pi } ^ { \pi } e ^ { \kappa \cos ( \theta - \theta_0 ) } \, \ud \theta.
\end{equation}
\par
The usual Gaussian distribution with mean $\mu$ and variance $\sigma ^ 2$ is denoted by $\mathcal{N} ( \mu, \sigma ^ 2 )$.

\subsubsection{Unbiased estimator}

For the estimator $\hat{\theta}$ of a quantity $\theta$, the bias is defined as the difference between the sampling expectation of $\hat{\theta}$ and $\theta$,
\begin{equation}
 \mathrm{bias} ( \hat{\theta} ) = \esp ( \hat{\theta} ) - \theta.
\end{equation}
If the bias is equal to 0, then the estimator is said to be unbiased. If the bias vanishes as the data size tends to infinity, then the estimator is said to be asymptotically unbiased.

\subsubsection{Bachmann--Landau notation}

Finally, a function $f ( N )$ is said to be $O ( 1 / N )$ if it is bounded by a function proportional to $1/N$, that is, for which there exists an $N_0$ and a $k > 0$ such that
$$\forall N > N_0 \quad \left| f ( N ) \right| < \frac{ k } { N }.$$

\subsection{Time-frequency analysis} \label{ss:tfa}

We here introduce the general framework for time-frequency analysis. More specifically, we consider $N$ signals $x_n ( t )$, $n = 1, \dots, N$, each signal being measured as the response to a  repetition of the stimulus. The $x_n ( t )$'s are assumed to be $N$ independent and identically distributed (i.i.d.) realizations of a signal $x ( t )$. Time-frequency analysis maps each $x_n ( t )$ into a complex time-frequency function $\TF{x_n}$ that we express as
\begin{equation} \label{eq:TF:ampl-phase}
 \TF{x_n} = | \TF{x_n} | \, e ^ { \phase{x_n} }.
\end{equation}
$| \TF{x_n} |$ is the amplitude of the \TFT\ and $| \TF{x_n} | ^ 2$ its power; $\phase{x_n}$ is the phase of the time-frequency transform. Since the $x_n ( t )$'s are i.i.d., so are their time-frequency transforms. As a consequence, the $\TF{x_n}$'s can be considered as i.i.d. realizations of a random  \TFT\ denoted $\TF{x}$ with amplitude $| \TF{x} |$ and phase $\phase{x}$.

\subsection{Three quantities of interest} \label{ss:tqoi}

We can now go back to our three quantities of interest: avgAMP, ITC and POWavg. To emphasize the fact that they depend on a a series of signals $x_1 ( t )$, \dots, $x_N ( t )$, as well as on the time $t$ and frequency $f$ at which the \TFT\ is computed, we write $\avgAMP{x}$, $\ITC{x}$ and $\POWavg{x}$. 

\subsubsection{avgAMP}

avgAMP is defined as in \eqref{eq:avgAMP:def}. It can be shown (see \S1.1 of \supplmat{1}) that the expectation of avgAMP\textsuperscript{2} can be expressed as
\begin{equation}  \label{eq:avgAMP2:esp}
 \esp \left[ \avgAMP{x} ^ 2 \right] = \esp \left[ \left| \TF{x} \right| \right] ^2 + \frac{ 1 } { N } \var \left[ \left| \TF{x} \right| \right].
\end{equation}
This first shows that $\esp ( \mathrm{avgAMP} ^ 2 )$ is only a function of $| \TF{x} |$, the amplitude of the \TFT---and therefore not of the phase $\phase{x}$. Also, from \eqref{eq:avgAMP2:esp}, we obtain that $\esp [ \avgAMP{x} ^ 2 ]$ tends to $\esp [ | \TF{x} | ] ^2$ as $N \to \infty$, i.e., avgAMP\textsuperscript{2} is an asymptotically unbiased estimator of $\esp [ | \TF{x} | ] ^2$. For finite $N$, its bias is given by $\var [ | \TF{x} | ] / N$. A last expression that can be derived from \eqref{eq:avgAMP2:esp} is the following asymptotic expression obtained by applying Bachmann--Landau notation:
\begin{equation} \label{eq:avgAMP2:asympt}
 \esp \left[ \avgAMP{x} ^ 2 \right] = \esp \left[ \left| \TF{x} \right| \right] ^2 + O \left( \frac{ 1 } { N } \right).
\end{equation}

\subsubsection{ITC}

ITC is defined as in \eqref{eq:ITC:def}. It can be shown (see \S1.2 of \supplmat{1}) that the expectation of ITC\textsuperscript{2} yields
\begin{equation} \label{eq:ITC2:esp}
 \esp \left[ \ITC{x} ^ 2 \right] = \left| \esp \left[ e ^ { i \phase{x} } \right] \right| ^ 2 + \frac{1}{N} \var \left[ e ^ { i \phase{x} } \right].
\end{equation}
This result shows that $\esp ( \mathrm{ITC} ^ 2 )$ is a only function of $\phase{x}$, the phase of the \TFT---and therefore not of the amplitude $| \TF{x} |$. It also shows that $\esp [ \ITC{x} ^ 2 ]$ tends to $| \esp [ e ^ { i \phase{x} } ] | ^ 2$ when $N \to \infty$, that is, ITC\textsuperscript{2} it is an asymptotically unbiased estimator of $| \esp [ e ^ { i \phase{x} } ] | ^ 2$, the squared mean resultant length of $\phase{x}$. For finite $N$, the bias is given by $\var [ e ^ { i \phase{x} } ] / N$. Finally, we obtain from \eqref{eq:ITC2:esp} the asymptotic expansion
\begin{equation} \label{eq:ITC2:asympt}
 \esp \left[ \ITC{x} ^ 2 \right] = \left| \esp \left[ e ^ { i \phase{x} } \right] \right| ^ 2 + O \left( \frac{ 1 } { N } \right).
\end{equation}

\subsubsection{POWavg}

POWavg is defined as in \eqref{eq:POWavg:def}, with
\begin{equation} \label{eq:POWavg:moy:def}
  \overline{ x_n } ( t ) = \frac{ 1 } { N } \sum_{ n = 1 } ^ N x_n ( t ).
\end{equation}
Due to the linearity of the time-frequency analysis, $\TF{\overline{x_n}}$ is equal to the average of the time-frequency transforms, 
\begin{equation} \label{eq:POWavg:tfsm}
 \TF{\overline{x_n}} = \frac{1}{N} \sum_{n = 1}^N \TF{x_n},
\end{equation}
so that
\begin{equation}
 \POWavg{x} = \left| \frac{1}{N} \sum_{n = 1} ^ N \TF{x_n} \right| ^ 2.
\end{equation}
It can be shown (see \S1.3 of \supplmat{1}) that the expectation of POWavg reads 
\begin{equation} \label{eq:POWavg:esp}
 \esp \left[ \POWavg{x} \right] = \left| \esp \left[ \TF{x} \right] \right| ^ 2 + \frac{1}{N} \var \left[ \TF{x} \right].
\end{equation}
Importantly, this result shows that, unlike avgAMP and ITC, POWavg depends on both the amplitude and the phase of the \TFT\ through the modulus and argument of $\TF{x}$---recall \eqref{eq:TF:ampl-phase}. From \eqref{eq:POWavg:esp}, we also have that $\esp [ \POWavg{x} ]$ tends to $| \esp [ \TF{x} ] | ^ 2$ as $N \to \infty$, leading to the conclusion that POWavg is an asymptotically unbiased estimator of $\left| \esp \left[ \TF{x} \right] \right| ^ 2$. For finite $N$, the bias is given by $\var [ \TF{x} ] / N$. Similarly to the two other measures, we asymptotically have
\begin{equation} \label{eq:POWavg:asympt}
 \esp \left[ \POWavg{x} \right] = \left| \esp \left[ \TF{x} \right] \right| ^ 2 + O \left( \frac{ 1 } { N } \right).
\end{equation}

\subsubsection{Summary}
 
Results regarding the general properties of the expectation of avgAMP\textsuperscript{2}, ITC\textsuperscript{2} and POWavg are summarized in \S1.4 of \supplmat{1}.

\subsection{Investigation of oscillatory model} \label{ss:ioom}

In this section, we investigate the behaviors of avgAMP, ITC, and POWavg in the particular case of an oscillatory signal.

\subsubsection{Model}

We assume that $x ( t )$ is a cosine with amplitude $\Omega$, frequency $\nu$, and phase $\phi$, i.e., of the form
\begin{equation} \label{eq:osc:m0:def}
 x ( t ) = \Omega \cos ( 2 \pi \nu t + \phi ).
\end{equation}
The parameters $\Omega$, $\nu$, and $\phi$ will either be assumed be constant, or follow a certain distribution over trials depending on the model (see \S2.6 of \supplmat{1} for a summary of models and results).

\subsubsection{S-transform}

The S-transform is defined as \gcite{Stockwell-1996}
\begin{equation} \label{eq:St:def}
 \TF{x} = \frac{ 1 } { \sqrt{ 2 \pi } } \int_{- \infty} ^{+ \infty} x ( u ) \, | f | \, e ^ { - \frac{ f ^ 2 ( u - t ) ^ 2 } { 2 } } \, e ^ { - 2 i \pi f u } \, \ud u.
\end{equation}
This is of the form
$$\frac{1}{\sqrt{2\pi\alpha^2}} \int_{- \infty} ^{+ \infty} x ( u ) \, e ^ { - \frac{ ( u - t ) ^ 2 } { 2 \alpha ^ 2 } } \, e ^ { - 2 i \pi f u } \, \ud u$$
with\footnote{Note that $\alpha$ is sometimes defined as $1 / ( f \sqrt{ 2 \pi } )$, so that $1 / \sqrt{ 2 \pi \alpha ^ 2 } = | f |$ and $1 / ( 2 \alpha ^2 ) = \pi f ^ 2$.} $\alpha = 1 / f$. The S-transform can therefore be seen as a band-pass filter or a windowed Fourier transform with a window whose width decreases with increasing frequency.

\subsubsection{\TFT\ of signal}

Direct calculation shows that the S-transform of $x ( t )$ is given by (see \S2.2 of \supplmat{1})
\begin{eqnarray}
 \TF{x} & = & \frac{ \Omega } { 2 } e ^ { - \frac{ 1 } { 2 } ( 2 \pi ) ^ 2 \left( 1 - \frac{ \nu } { f } \right) ^ 2 } e ^ { i [ \phi - 2 \pi ( f - \nu ) t ] } \nonumber \\
 & & \qquad + \frac{ \Omega } { 2 } e ^ { - \frac{ 1 }{ 2 } ( 2 \pi ) ^ 2 \left( 1 + \frac{ \nu } { f } \right) ^ 2 } e ^ { i [ - \phi - 2 \pi ( f + \nu ) t ] }. \nonumber \\ \label{eq:osc:gen:tf}
\end{eqnarray}
The amplitude and phase of this complex number do not have simple exact forms. Still, it can be shown that a very good approximation of $\TF{x}$ can be obtained by only keeping the first term of the right-hand side of \eqref{eq:osc:gen:tf}. More precisely, we have (see \S2.3 of \supplmat{1})
\begin{equation} \label{eq:osc:gen:approx:rel}
 \TF{x} = \TFa{x} \left[ 1 + \epsilon ( t, f ) \right]
\end{equation}
with 
\begin{equation} \label{eq:osc:gen:approx:TFa}
 \TFa{x} = \frac{ \Omega } { 2 } e ^ { - \frac{ 1 } { 2 } ( 2 \pi ) ^ 2 \left( 1 - \frac{ \nu } { f } \right) ^ 2 } e ^ { i [ \phi - 2 \pi ( f - \nu ) t ] }
\end{equation}
and
\begin{equation} \label{eq:osc:gen:approx:eps}
 \left| \epsilon ( t, f ) \right| = e ^ { - 8 \pi ^ 2 \frac{ \nu } { f } }.
\end{equation}
The modulus and argument of $\TFa{x}$ are given by 
\begin{equation} \label{eq:osc:gen:approx:TFa:mod}
 \left| \TFa{x} \right | =  \frac{ \Omega } { 2 } e ^ { - \frac{ 1 } { 2 } ( 2 \pi ) ^ 2 \left( 1 - \frac{ \nu } { f } \right) ^ 2 }
\end{equation}
and
\begin{equation} \label{eq:osc:gen:approx:TFa:arg}
 \arg \left[ \TFa{x} \right] = \phi - 2 \pi ( f - \nu ) t,
\end{equation}
respectively. Since we are mostly interested in frequencies in the range $0 < f < 10 \nu$, the relative error $\epsilon ( t, f )$ is bounded by $| \epsilon ( t, f ) | < 3.8 \times 10 ^ { - 4 }$ (see \S2.3 of \supplmat{1}). The amplitude of $\TF{x}$ is given by
\begin{equation} \label{eq:osc:gen:approx:ampl}
 \left| \TF{x} \right| = \frac{ \Omega } { 2 } e ^ { - \frac{ 1 } { 2 } ( 2 \pi ) ^ 2 \left( 1 - \frac{ \nu } { f } \right) ^ 2 } \left[ 1 + \epsilon_m ( t, f ) \right]
\end{equation}
and the phase of $\TF{x}$ by
\begin{equation} \label{eq:osc:gen:approx:phase}
 \phase{x} \approx \phi - 2 \pi ( f - \nu ) t + \epsilon_a ( t, f ).
\end{equation}
For more detail on the errors $\epsilon_m ( t, f )$ and $\epsilon_a ( t, f )$, see \S2.3 of \supplmat{1}.
\par
As expected, there is a strong connection between the amplitude and phase of the signal and those of the \TFT: the amplitude of the \TFT, \eqref{eq:osc:gen:approx:ampl}, only depends on the amplitude (and frequency) of the signal, while the phase of the \TFT, \eqref{eq:osc:gen:approx:phase}, only depends on the phase (and frequency) of the signal.
\par
Note that the amplitude of the \TFT\ is not a function of $t$. As a function of $f$, it is maximal for $f = \nu$ (and equal to $\Omega / 2$) and decreases when $f$ moves away from $\nu$. The phase of the \TFT\ is a (bilinear) function of both $t$ and $f$ that is equal to $\phi$ for pairs $( t , f )$ of the form either $( 0, f )$ for any $f$ or $( t, \nu )$ for  any $t$.

\subsubsection{Simplest model of repetition} \label{sss:m1}

We now consider $N$ repetitions of the oscillatory signal of \eqref{eq:osc:m0:def} with constant amplitude ($\Omega = \Omega_0$) and frequency ($\nu = \nu_0$) but varying phases $\phi_n$, $n = 1, \dots, N$,
\begin{equation} \label{eq:osc:m1:def}
 x_n ( t ) = \Omega_0 \cos ( 2 \pi \nu_0 t + \phi_n ), \qquad n = 1, \dots, N.
\end{equation}
We assume that the $\phi_n$'s are i.i.d. repetitions of $\phi \sim \mathrm{VonMises} (\phi_0,\kappa)$. We define the circular mean as in \eqref{eq:moycirc:def}
\begin{equation} \label{eq:osc:moycirc}
 \esp ( e ^ { i \phi } ) = \rho e ^ { i \phi_0 }, 
\end{equation}
with $\phi_0$ the mean direction and $\rho$ the mean resultant length. Applying the approximation of \eqref{eq:osc:gen:approx:TFa}, the \TFT\ of each signal $x_n(t)$ is given by
\begin{equation} \label{eq:osc:m1:tf}
 \TFa{x_n} = \frac{ \Omega_0 } { 2 } e ^ { - \frac{ 1 } { 2 } ( 2 \pi ) ^ 2 \left( 1 - \frac{ \nu_0 } { f } \right) ^ 2 } e ^ { i [ \phi_n - 2 \pi ( f - \nu_0 ) t ] }. 
\end{equation}
The amplitude is equal to 
\begin{equation} \label{eq:osc:m1:ampl}
 \left| \TFa{x_n} \right| = \frac{ \Omega_0 } { 2 } e ^ { - \frac{ 1 } { 2 } ( 2 \pi ) ^ 2 \left( 1 - \frac{ \nu_0 } { f } \right) ^ 2 },
\end{equation}
and the phase to
\begin{equation} \label{eq:osc:m1:phas}
 \arg \left[ \TFa{x_n} \right] = \phi_n - 2 \pi ( f - \nu_0 ) t.
\end{equation}
In this model, the phase depends on $n$ and is random (as it depends on $\phi_n$), whereas the amplitude does not depend on $n$ and is deterministic. From \eqref{eq:avgAMP:def} and \eqref{eq:osc:m1:ampl}, we obtain
\begin{equation} \label{eq:osc:m1:avgAMP}
 \avgAMPa{x} = \frac{ \Omega_0 } { 2 } e ^ { - \frac{ 1 } { 2 } ( 2 \pi ) ^ 2 \left( 1 - \frac{ \nu_0 } { f } \right) ^ 2 },
\end{equation}
and, from \eqref{eq:ITC:def} and \eqref{eq:osc:m1:phas},
\begin{equation} \label{eq:osc:m1:ITC}
 \ITCa{x} = \left| \frac{1}{N} \sum_{ n = 1 }^ N e ^ { i \phi_n } \right|.
\end{equation}
Besides, the \TFT\ of the average signal can be calculed using \eqref{eq:POWavg:tfsm} and \eqref{eq:osc:m1:tf}, yielding
\begin{equation}
 \TFa{\overline{ x_n }} = \frac{ \Omega_0 } { 2 } e ^ { - \frac{ 1 } { 2 } ( 2 \pi ) ^ 2 \left( 1 - \frac{ \nu_0 } { f } \right) ^ 2 } \, e ^ { - 2 i \pi ( f - \nu_0 ) t } \, \frac{1}{N} \sum_{ n = 1 }^ N e ^ { i \phi_n },
\end{equation}
which has amplitude
\begin{equation}
 \left| \TFa{\overline{ x_n }} \right| = \frac{ \Omega_0 } { 2 } e ^ { - \frac{ 1 } { 2 } ( 2 \pi ) ^ 2 \left( 1 - \frac{ \nu_0 } { f } \right) ^ 2 } \left| \frac{1}{N} \sum_{ n = 1 }^ N e ^ { i \phi_n } \right|
\end{equation}
(the phase is irrelevant for our calculation). This, together with \eqref{eq:POWavg:def}, leads to
\begin{equation} \label{eq:osc:m1:POWavg}
 \POWavga{x} = \left[ \frac{ \Omega_0 } { 2 } e ^ { - \frac{ 1 } { 2 } ( 2 \pi ) ^ 2 \left( 1 - \frac{ \nu_0 } { f } \right) ^ 2 } \right] ^ 2 \left| \frac{1}{N} \sum_{ n = 1 }^ N e ^ { i \phi_n } \right| ^ 2.
\end{equation}
From \eqref{eq:osc:m1:avgAMP}, \eqref{eq:osc:m1:ITC}, and \eqref{eq:osc:m1:POWavg}, it is straightforward to see that we have the following relationship
\begin{equation} \label{eq:osc:m1:rel}
 \mathrm{POWavg}\approxim = \left( \mathrm{avgAMP}\approxim \right) ^ 2 \times \left( \mathrm{ITC}\approxim \right) ^ 2.
\end{equation}

\subsubsection{Model with varying amplitude and phase} \label{sss:m2}

We now consider a more complex model, in which we observe repetitions of an oscillatory signal with constant frequency ($\nu = \nu_0$) but varying amplitude and phase,
\begin{equation} \label{eq:osc:m2:def}
 x_n ( t ) = \Omega_n \cos ( 2 \pi \nu_0 t + \phi_n ), \qquad n = 1, \dots, N. 
\end{equation}
We assume that the $\Omega_n$'s are i.i.d. repetitions of $\Omega \sim \mathcal{N} ( \Omega_0, \tau_{\Omega}^2 )$, while the $\phi_n$'s are i.i.d repetitions of $\phi \sim \mathrm{VonMises} (\phi_0,\kappa)$. Furthermore, $\Omega_n$ and $\phi_n$ are assumed to be independent from each other for every $n$.
\par
According to \eqref{eq:osc:gen:approx:TFa}, the (approximate) S-transform of each signal is given by
\begin{equation} \label{eq:osc:m2:tf}
 \TFa{x_n} = \frac{ \Omega_n } { 2 } e ^ { - \frac{ 1 } { 2 } ( 2 \pi ) ^ 2 \left( 1 - \frac{ \nu_0 } { f } \right) ^ 2 } e ^ { i [ \phi_n - 2 \pi ( f - \nu_0 ) t ] },
\end{equation}
which has amplitude 
\begin{equation} \label{eq:osc:m2:ampl}
 \left| \TFa{x_n} \right| = \frac{ \Omega_n } { 2 } e ^ { - \frac{ 1 } { 2 } ( 2 \pi ) ^ 2 \left( 1 - \frac{ \nu_0 } { f } \right) ^ 2 } 
\end{equation}
and phase
\begin{equation} \label{eq:osc:m2:phas}
  \arg \left[ \TFa{x_n} \right] = \phi_n - 2 \pi ( f - \nu_0 ) t.
\end{equation}
In this case, both amplitude and phase vary with $n$ and are random (through their dependence in $\Omega_n$ for amplitude, and $\phi_n$ for phase). Also, since $\Omega_n$ and $\phi_n$ are independent for every $n$, so are $| \TFa{x_n} |$ and $\arg [ \TFa{x_n} ]$.
\par
From \eqref{eq:avgAMP:def} and \eqref{eq:osc:m2:ampl}, we have
\begin{equation}
 \avgAMPa{x} = \frac{ 1 } { 2 } e ^ { - \frac{ 1 } { 2 } ( 2 \pi ) ^ 2 \left( 1 - \frac{ \nu_0 } { f } \right) ^ 2 }  \frac{ 1 } { N } \sum_{ n = 1 } ^ N \Omega_n,
\end{equation}
and, from \eqref{eq:ITC:def} and \eqref{eq:osc:m2:phas},
\begin{equation}
 \ITCa{x} = \left| \frac{1}{N} \sum_{ n = 1 }^ N  e ^ { i \phi_n } \right|,
\end{equation}
so that
\begin{eqnarray}
 & & \avgAMPa{x} ^ 2 \times \ITCa{x} ^ 2 \nonumber \\
 & = & \left[ \frac{ 1 } { 2 } e ^ { - \frac{ 1 } { 2 } ( 2 \pi ) ^ 2 \left( 1 - \frac{ \nu_0 } { f } \right) ^ 2 } \right] ^ 2 \left( \frac{ 1 } { N } \sum_{ n = 1 } ^ N \Omega_n \right) ^ 2 \left| \frac{1}{N} \sum_{ n = 1 }^ N  e ^ { i \phi_n } \right| ^ 2. \nonumber \\ \label{eq:osc:m2:ITC-avgAMP}
\end{eqnarray}
Besides, the \TFT\ of the average signal is equal to the average of the $\TF{x_n}$'s from Equation~(\ref{eq:osc:m2:tf}), that is,
\begin{equation} \label{eq:osc:m2:sm:tf}
 \TFa{ \overline{ x_n } } = \frac{ 1 } { 2 } e ^ { - \frac{ 1 } { 2 } ( 2 \pi ) ^ 2 \left( 1 - \frac{ \nu_0 } { f } \right) ^ 2 }  e ^ { - 2 i \pi ( f - \nu_0 ) t } \, \frac{1}{N} \sum_{ n = 1 }^ N  \Omega_n \, e ^ { i \phi_n }.
\end{equation}
The amplitude of this quantity is given by
\begin{equation} \label{eq:osc:m2:AMPavg}
 \left| \TFa{ \overline{ x_n } } \right| = \frac{ 1 } { 2 } e ^ { - \frac{ 1 } { 2 } ( 2 \pi ) ^ 2 \left( 1 - \frac{ \nu_0 } { f } \right) ^ 2 } \left| \frac{1}{N} \sum_{ n = 1 }^ N  \Omega_n e ^ { i \phi_n } \right|.
\end{equation}
Incorporating this result into \eqref{eq:POWavg:def}, we are led to
\begin{equation} \label{eq:osc:m2:POWavg}
 \POWavga{x} = \left[ \frac{ 1 } { 2 } e ^ { - \frac{ 1 } { 2 } ( 2 \pi ) ^ 2 \left( 1 - \frac{ \nu_0 } { f } \right) ^ 2 } \right] ^ 2 \left| \frac{1}{N} \sum_{ n = 1 }^ N  \Omega_n e ^ { i \phi_n } \right| ^ 2.
\end{equation}
Comparing \eqref{eq:osc:m2:ITC-avgAMP} and \eqref{eq:osc:m2:POWavg}, we see that, in this particular case, the relationship of \eqref{eq:osc:m1:rel} does not hold. Indeed, the influences of the signals' amplitudes (the $\Omega_n$'s) and phases (the $\phi_n$'s) are separated in \eqref{eq:osc:m2:ITC-avgAMP} but appear jointly in \eqref{eq:osc:m2:POWavg}.
\par
Still, one can take advantage of the statistical independence of the $\Omega_n$'s and the $\phi_n$'s to derive the expectations of the various quantities. We start by calculating the expectation of amplitude from \eqref{eq:osc:m2:ampl}, leading to
\begin{equation} \label{eq:osc:m2:ampl:esp}
 \esp \left[ \left| \TFa{x_n} \right| \right] = \frac{ \Omega_0 } { 2 } e ^ { - \frac{ 1 } { 2 } ( 2 \pi ) ^ 2 \left( 1 - \frac{ \nu_0 } { f } \right) ^ 2 }.
\end{equation}
An asymptotic approximation of $\esp [ ( \mathrm{avgAMP}\approxim ) ^ 2 ]$ is then available from \eqref{eq:avgAMP2:asympt} and \eqref{eq:osc:m2:ampl:esp}
\begin{equation} \label{eq:osc:m2:avgAMP2:esp}
 \esp \left[ \avgAMPa{x} ^ 2 \right] = \left[ \frac{ \Omega_0 } { 2 } e ^ { - \frac{ 1 } { 2 } ( 2 \pi ) ^ 2 \left( 1 - \frac{ \nu_0 } { f } \right) ^ 2 } \right] ^ 2 + O \left( \frac{1}{N} \right).
\end{equation}
One can also compute the circular mean by using \eqref{eq:moycirc:def} and \eqref{eq:osc:m2:phas},
\begin{equation} \label{eq:osc:m2:phas:esp}
  \esp \left[ e ^ { i \arg \left[ \TFa{x_n} \right] } \right] = \rho e ^ { i [ \phi_0 - 2 \pi ( f - \nu_0 ) t ] },
\end{equation}
and, from \eqref{eq:ITC2:asympt},
\begin{equation} \label{eq:osc:m2:ITC2:esp}
 \esp \left[ \ITCa{x} ^ 2 \right] = \rho ^ 2 + O \left( \frac{ 1 } { N } \right).
\end{equation}
The product $\esp [ ( \mathrm{avgAMPL}\approxim ) ^ 2 ]$ by $\esp [ ( \mathrm{ITC}\approxim ) ^ 2 ]$ can then be calculated from \eqref{eq:osc:m2:avgAMP2:esp} and \eqref{eq:osc:m2:ITC2:esp}, yielding
\begin{eqnarray*}
 & & \esp \left[ \avgAMPa{x} ^ 2 \right] \times \esp \left[ \ITCa{x} ^ 2 \right] \\
 & = & \left[ \frac{ \Omega_0 \rho } { 2 } e ^ { - \frac{ 1 } { 2 } ( 2 \pi ) ^ 2 \left( 1 - \frac{ \nu_0 } { f } \right) ^ 2 } \right] ^ 2 + O \left( \frac{1}{N} \right).
\end{eqnarray*}
Since $| \TF{x} |$ and $\arg [ \TFa{x} ]$ are independent, so are avgAMP$\approxim$ and ITC$\approxim$, so that, according to \eqref{eq:esp:prop},
\begin{eqnarray} \label{eq:osc:m2:prod}
 & & \esp \left[ \ITCa{x} ^ 2 \times \avgAMPa{x} ^ 2 \right] \nonumber \\
 & = & \left[ \frac{ \Omega_0 \rho } { 2 } e ^ { - \frac{ 1 } { 2 } ( 2 \pi ) ^ 2 \left( 1 - \frac{ \nu_0 } { f } \right) ^ 2 } \right] ^ 2 + O \left( \frac{1}{N} \right).
\end{eqnarray}
Deriving an asymptotic form for $\esp ( \mathrm{POWavg}\approxim )$ from \eqref{eq:POWavg:asympt} and \eqref{eq:osc:m2:sm:tf} is lengthier, but it can be done and leads to (see \S2.4 of \supplmat{1})
\begin{equation} \label{eq:osc:m2:AMPavg:esp}
 \esp [ \POWavga{x} ] = \left[ \frac{ \Omega_0 \rho } { 2 } e ^ { - \frac{ 1 } { 2 } ( 2 \pi ) ^ 2 \left( 1 - \frac{ \nu_0 } { f } \right) ^ 2 } \right] ^ 2 + O \left( \frac{ 1 } { N } \right).
\end{equation}
Equality of \eqref{eq:osc:m2:prod} and \eqref{eq:osc:m2:AMPavg:esp} shows that we have
\begin{equation}
 \esp \left( \mathrm{POWavg}\approxim \right) = \esp \left[ \left( \mathrm{avgAMP}\approxim \right) ^ 2 \times \left( \mathrm{ITC}\approxim \right) ^ 2 \right] + O \left( \frac{1}{N} \right). 
\end{equation}
This result is weaker than in the simplest model leading to \eqref{eq:osc:m1:rel}, in that (i) it involves an equality in expectation, and (ii) it is only true asymptotically.

\subsubsection{Model with varying amplitude, frequency, and phase} \label{sss:m3}

We finally consider repetitions of the oscillatory signal with varying amplitude, phase, and frequency
\begin{equation}
 x_n ( t ) = \Omega_n \cos ( 2 \pi \nu_n t + \phi_n ), \qquad n = 1, \dots, N. 
\end{equation}
We assume that we have $\Omega_n \sim \mathcal{N} ( \Omega_0, \tau_{\Omega}^2 )$, $\nu_n \sim \mathcal{N} ( \nu_0, \tau_{\nu}^2 )$, and $\phi_n \sim \mathrm{VonMises} (\phi_0,\kappa)$. 

The approximate \TFT\ of each signal is given by
\begin{equation} \label{eq:osc:m3:tf}
 \TFa{x_n} = \frac{ \Omega_n } { 2 } e ^ { - \frac{ 1 } { 2 } ( 2 \pi ) ^ 2 \left( 1 - \frac{ \nu_n } { f } \right) ^ 2 } e ^ { i [ \phi_n - 2 \pi ( f - \nu_n ) t ] },
\end{equation}
with amplitude 
\begin{equation} \label{eq:osc:m3:ampl}
  \left| \TFa{x_n} \right| = \frac{ \Omega_n } { 2 } e ^ { - \frac{ 1 } { 2 } ( 2 \pi ) ^ 2 \left( 1 - \frac{ \nu_n } { f } \right) ^ 2 } 
\end{equation}
and phase
\begin{equation} \label{eq:osc:m3:phas}
 \arg \left[ \TFa{x_n} \right] = \phi_n - 2 \pi ( f - \nu_n ) t.
\end{equation}
Note that, in this case, amplitude and phase are \emph{not} independent, as they both depend on $\nu_n$. As in the previous model, we will need to calculate expectations. From Equation~(\ref{eq:osc:m3:ampl}), we obtain (see \S2.5.2 of \supplmat{1})
\begin{equation}
 \esp \left[ \left| \TFa{x_n} \right| \right] = \frac{ \Omega_0 } { 2 }  \frac{ 1 } { \sqrt{ \left( \frac{ 2 \pi \tau_{\nu} } { f } \right) ^ 2 + 1 } } e ^ { - \frac{ 1 } { 2 } \frac{ ( 2 \pi ) ^ 2 } { \left( \frac{ 2 \pi \tau_{\nu} } { f } \right) ^ 2 + 1 } \left( 1 - \frac{ \nu_0 } { f } \right) ^2 }
\end{equation}
and, from \eqref{eq:avgAMP2:asympt},
\begin{eqnarray}
 & & \esp \left[ \avgAMPa{x} ^ 2 \right] \\
 & = & \left[ \frac{ \Omega_0 } { 2 }  \frac{ 1 } { \sqrt{ \left( \frac{ 2 \pi \tau_{\nu} } { f } \right) ^ 2 + 1 } } e ^ { - \frac{ 1 } { 2 } \frac{ ( 2 \pi ) ^ 2 } { \left( \frac{ 2 \pi \tau_{\nu} } { f } \right) ^ 2 + 1 } \left( 1 - \frac{ \nu_0 } { f } \right) ^2 } \right] ^ 2 \nonumber \\
 & & \quad + O \left( \frac{1}{N} \right). \label{eq:osc:m3:avgAMP2:esp}
\end{eqnarray}
From \eqref{eq:osc:m3:phas}, we obtain (see \S2.5.3 of \supplmat{1})
\begin{equation}
 \esp \left\{ e ^ { i \arg \left[ \TFa{x_n} \right] } \right\} = \rho e ^ { - \frac{ 1 } { 2 } ( 2 \pi t ) ^ 2 \tau_{\nu} ^ 2 } e ^ { i \left[ \phi_0 - 2 \pi ( f - \nu_0 ) t \right] },
\end{equation}
so that
\begin{equation}
 \left| \esp \left\{ e ^ { i \arg \left[ \TFa{x_n} \right] } \right\} \right| = \rho e ^ { - \frac{ 1 } { 2 } ( 2 \pi t ) ^ 2 \tau_{\nu} ^ 2 },
\end{equation}
and, from \eqref{eq:ITC2:asympt},
\begin{equation} \label{eq:osc:m3:ITC2:esp}
 \esp \left[ \ITCa{x} ^ 2 \right] = \rho ^ 2 \left[ e ^ { - \frac{ 1 } { 2 } ( 2 \pi t ) ^ 2 \tau_{\nu} ^ 2 } \right] ^ 2 + O \left( \frac{ 1 } { N } \right).
\end{equation}
To calculate POWavg$\approxim$, we need the following quantity (see \S2.5.4 of \supplmat{1}):
\begin{eqnarray*}
 \esp \left[ \TFa{x_n} \right] & = & \frac{ 1 } { 2 } \esp ( \Omega_n ) \esp \left( e ^ { i \phi_n } \right) \nonumber \\
 & & \quad \times  \esp \left[ e ^ { - \frac{ 1 } { 2 } ( 2 \pi ) ^ 2 \left( 1 - \frac{ \nu_n } { f } \right) ^ 2 - 2 i \pi ( f - \nu_n ) t } \right] \\
 & = &  \frac{ \Omega_0 \rho } { 2 } \frac{ 1 } { \sqrt{ \left( \frac{ 2 \pi \tau_{\nu} } { f } \right) ^ 2 + 1 } } e ^ { - 2 i \pi ( f - \nu_0 ) t } \\
 & & \qquad \times e ^ { - \frac{ 1 } { 2 } \frac{ ( 2 \pi ) ^ 2 } { \left( \frac{ 2 \pi \tau_{\nu} } { f } \right) ^ 2 + 1 } \left[ \left( 1 - \frac{ \nu_0 } { f } \right) ^ 2 + \tau_{\nu} ^ 2 t ^ 2 \right] }.
\end{eqnarray*}
Finally, from \eqref{eq:POWavg:asympt}, 
\begin{eqnarray}
 & & \esp \left[ \POWavga{x} \right] \nonumber \\
 & = & \left[ \frac{ \Omega_0 } { 2 }  \frac{ 1 } { \sqrt{ \left( \frac{ 2 \pi \tau_{\nu} } { f } \right) ^ 2 + 1 } } e ^ { - \frac{ 1 } { 2 } \frac{ ( 2 \pi ) ^ 2 } { \left( \frac{ 2 \pi \tau_{\nu} } { f } \right) ^ 2 + 1 } \left( 1 - \frac{ \nu_0 } { f } \right) ^2 } \right] ^ 2 \nonumber \\
 & & \quad \times \rho ^ 2 \left[ e ^ { - \frac{ 1 } { 2 } \frac{ ( 2 \pi ) ^ 2 } { \left( \frac{ 2 \pi \tau_{\nu} } { f } \right) ^ 2 + 1 } \tau_{\nu} ^ 2 t ^ 2 } \right] ^ 2 + O \left( \frac{1}{N} \right). \label{eq:osc:m3:POWavg:esp}
\end{eqnarray}
From \eqref{eq:osc:m3:avgAMP2:esp}, \eqref{eq:osc:m3:ITC2:esp}, and \eqref{eq:osc:m3:POWavg:esp}, we are now in position to calculate
\begin{eqnarray} \label{eq:osc:m3:rel}
 & & \esp \left[ \POWavga{x} \right] \nonumber \\
 & & \qquad - \esp \left[ \ITCa{x} ^ 2 \times \avgAMPa{x} ^ 2 \right] \nonumber \\
 & = & \left[ \frac{ \Omega_0 \rho } { 2 }  \frac{ 1 } { \sqrt{ \left( \frac{ 2 \pi \tau_{\nu} } { f } \right) ^ 2 + 1 } } e ^ { - \frac{ 1 } { 2 } \frac{ ( 2 \pi ) ^ 2 } { \left( \frac{ 2 \pi \tau_{\nu} } { f } \right) ^ 2 + 1 } \left( 1 - \frac{ \nu_0 } { f } \right) ^2 } \right] ^ 2 \nonumber \\
 & & \ \times \left\{ \left[ e ^ { - \frac{ 1 } { 2 } \frac{ ( 2 \pi ) ^ 2 } { \left( \frac{ 2 \pi \tau_{\nu} } { f } \right) ^ 2 + 1 } \tau_{\nu} ^ 2 t ^ 2 } \right] ^ 2 - 1 \right\} + O \left( \frac{1}{N} \right). 
\end{eqnarray}
This difference, which is negative, does not tend to 0 as  $N \to \infty$. However, it becomes increasingly smaller when $\tau_{\nu}$ becomes smaller and vanishes for $\tau_{\nu} = 0$, which corresponds to a situation where the signal frequency is fixed, leading to independent amplitude and phase for the \TFT.

\subsection{General expression} \label{ss:ge}

In this section, we confirm the results of the previous section and show in a more general setting that
\begin{equation} \label{eq:rel:esp}
 \esp \left[ \mathrm{POWavg} - \mathrm{ITC} ^ 2 \times \mathrm{avgAMP} ^ 2 \right] = O \left( \frac{1}{N} \right),
\end{equation}
provided that a certain condition is met, which includes the case where the covariance between the amplitude $| \TF{x} |$ and phase $\phase{x}$ of $\TF{x}$ is equal to 0. Since independence entails a covariance of 0, independence is a sufficient condition.

\subsubsection{Case of independent amplitude and phase} \label{ss:ifc}

Given the asymptotic expansions of avgAMP, ITC, and POWavg, we can expect the following behaviors for large $N$. On the one hand, from \eqref{eq:avgAMP2:asympt} and \eqref{eq:ITC2:asympt}, we obtain
\begin{eqnarray} \label{eq:approx:prod}
 & & \avgAMP{x} ^ 2 \times \ITC{x} ^ 2  \nonumber \\
 & = & \esp \left[ \left| \TF{x} \right| \right] ^2 \, \left| \esp \left[ e ^ { i \phase{x} } \right] \right| ^ 2 + O \left( \frac{1}{N} \right)
\end{eqnarray}
On the other hand, starting from \eqref{eq:POWavg:asympt} and, remembering the expression of $\TF{x}$ in terms of $| \TF{x} |$ and $e ^ { i \phase{x} }$, \eqref{eq:TF:ampl-phase}, we have
\begin{eqnarray}
 & & \esp \left[ \POWavg{x} \right] \nonumber \\
 & = & \left| \esp \left[ | \TF{x} | e ^ { i \phase{x} } \right] \right| ^ 2 + O \left( \frac{ 1 } { N } \right).
\end{eqnarray}
Now, if we assume independence of $| \TF{x} |$ and $e ^ { i \phase{x} }$, we can apply \eqref{eq:esp:prop}, leading to
\begin{eqnarray} \label{eq:approx:POWavg:indep}
 & & \esp \left[ \POWavg{x} \right] \nonumber \\
 & = & \esp \left[ \left| \TF{x}\right| \right] ^ 2 \, \left| \esp \left[ e ^ { i \phase{x} } \right] \right| ^ 2 + O \left( \frac{ 1 } { N } \right).
\end{eqnarray}
Since \eqref{eq:approx:prod} and \eqref{eq:approx:POWavg:indep} are identical, we showed that 
\begin{equation} \label{eq:approx:hyp}
 \esp ( \mathrm{POWavg} ) = \esp ( \mathrm{avgAMP} ^ 2 \times \mathrm{ITC} ^ 2 ) + O \left( \frac{ 1 } { N } \right)
\end{equation}
in the case of independent amplitude and phase of the \TFT.

\subsubsection{General case}

We now proceed to a more general proof, showing in the process that having a covariance of 0 is a sufficient condition. To this end, we first calculate avgAMP\textsuperscript{2} $\times$ ITC\textsuperscript{2} from \eqref{eq:avgAMP:def} and \eqref{eq:ITC:def}, expanding the product by brute force, and then calculate its expectation term by term. We then calculate POWavg and its expectation using \eqref{eq:POWavg:asympt}, \eqref{eq:TF:ampl-phase}, and \eqref{eq:def:cov}. Calculating the difference between the two terms leads to (see \S3 of \supplmat{1}):
\begin{eqnarray} \label{eq:rel:res}
 & & \esp ( \mathrm{POWavg} - \mathrm{ITC} ^ 2 \times \mathrm{avgAMP} ^ 2 ) \nonumber \\
 & = & \left| \esp \left[ e ^ { i \phase{x} } \right] \esp \left[ \left| \TF{x} \right| \right] + \cov \left[ e ^ { i \phase{x} }, \left| \TF{x} \right| \right] \right| ^ 2 \nonumber \\
 & & \quad - \left| \esp \left[ e ^ { i \phase{x} } \right] \right| ^ 2 \esp \left[ \left| \TF{x} \right| \right] ^ 2 + O \left( \frac{1}{N} \right).
\end{eqnarray}
If $\cov \left[ e ^ { i \phase{x} }, \left| \TF{x} \right| \right] = 0$, then the right-hand side of the equation is equal to $O ( 1 / N )$. A sufficient condition for this covariance to be equal to 0 is for $e ^ { i \phase{x} }$ and $\left| \TF{x} \right|$ to be statistically independent. But this is not a necessary condition, as we can have non-independent (non-Gaussian) variables whose covariance is equal to 0. If the covariance is different from 0, it is still possible for the right-hand side of \eqref{eq:rel:res} to be $O ( 1 / N )$, although the corresponding solutions cannot be associated with simple cases.

\section{Simulation study} \label{s:ss}

In this section, we investigated the connection between avgAMP, ITC, and POWavg by generating synthetic data in the context of high-frequency oscillations (HFOs). Application of time-frequency analysis to brain recordings from  event-related protocols involving sensory stimulation have indicated that phase locking of ongoing EEG activity strongly contributes to HFOs (range 400--800~Hz) which are superimposed on the somatosensory evoked potential (SEP) \gcite{Curio-1994b, Curio-1997, Ozaki-1998, Valencia-2006}. The corresponding detailed temporal components of short durations and small amplitudes were generally identified by measuring latencies of the averaged evoked potential in time domain \gcite{Ozaki-1998}. In addition, ITC analysis was found to mimic the time-frequency power distribution of the HFOs, while the average of the single-sweep time-frequency power transforms did not show any significant increment of power \gcite{Valencia-2006}.

\subsection{Data generation}

We generated signals on a time window of $[ -100, 100]$~ms at a sampling rate of $f_s = 2$~kHz (corresponding to a recording every $\delta t = 0.5$~ms). Signals corresponding to the different trials were generated independently. For each trial $n$, we simulated an induced response in the $[20, 30]$~ms time window, and ongoing activity the rest of the time. Both the ongoing and the induced activities were generated using \eqref{eq:osc:m0:def} with the same amplitude $\Omega_n$ and frequency $\nu_n$, but with different phase: $\phi_n^{(o)}$ for the ongoing activity, and $\phi_n^{(i)}$ for the induced activity. The exact values of parameters were sampled according to specific distributions (see Table~\ref{tab:simu:param}). To limit the effect of spectral leakage on ITC, we added Gaussian white noise with a standard deviation of 0.01 (corresponding to 1\% of the expected signal amplitude). The signals were analyzed using the $S$-transform, as well as two wavelet transforms (Morse wavelet and analytic Morlet wavelet). Only results of the S-transform applied to the signals with $\kappa ^{(i)} = 10$ are reported here; see \S1 of \supplmat{2} for all results.

\begin{table}[!htbp]
 \centering
 \caption{\textbf{Simulation study.} Values of parameters for data generation.} \label{tab:simu:param}
 \begin{tabular}{c|c|cc}
   Parameter & Distribution & \multicolumn{2}{c}{Parameters} \\
   \hline
   $\Omega_n$ & $\mathcal{N} ( \Omega_0, \tau_{\Omega} ^ 2 )$ & $\Omega_0 = 1$ & $\tau_{ \Omega } = 0.1$ \\
   $\nu_n$ & $\mathcal{N} ( \nu_0, \tau_{\nu} ^ 2 )$ & $\nu_0 = 500$ & $\tau_{ \nu } \in \{ 0, 5 \}$ \\
   $\phi_n^{(o)}$ & $\mathrm{vonMises} [ \phi_0, \kappa^{(o)} ]$ & $\phi_0 = 0$ & $\kappa^{(o)} = 0$ \\
   $\phi_n^{(i)}$ & $\mathrm{vonMises} [ \phi_0, \kappa^{(i)} ]$ & $\phi_0 = 0$ & $\kappa^{(i)} \in \{ 1, 10 \}$
 \end{tabular}
\end{table}

\subsection{Assessment analysis}

Based on the simulated data, we computed avgAMP\textsuperscript{2}, ITC\textsuperscript{2}, POWavg, as well as the difference between POWavg and $\mathrm{avgAMP}^2 \times \mathrm{ITC}^2$. To assess the validity of the assumption of a covariance of 0 between the amplitude and phase of the \TFT, we also computed the module of the sample covariance between $e ^ { i \phase{x} }$ and $\left| \TF{x} \right|$.

\subsection{Results}

\begin{figure*}[!htbp]
  \centering
  \begin{tabular}{cc}
  POWavg & avgAMP\textsuperscript{2} \\
  \includegraphics[width=0.5\linewidth]{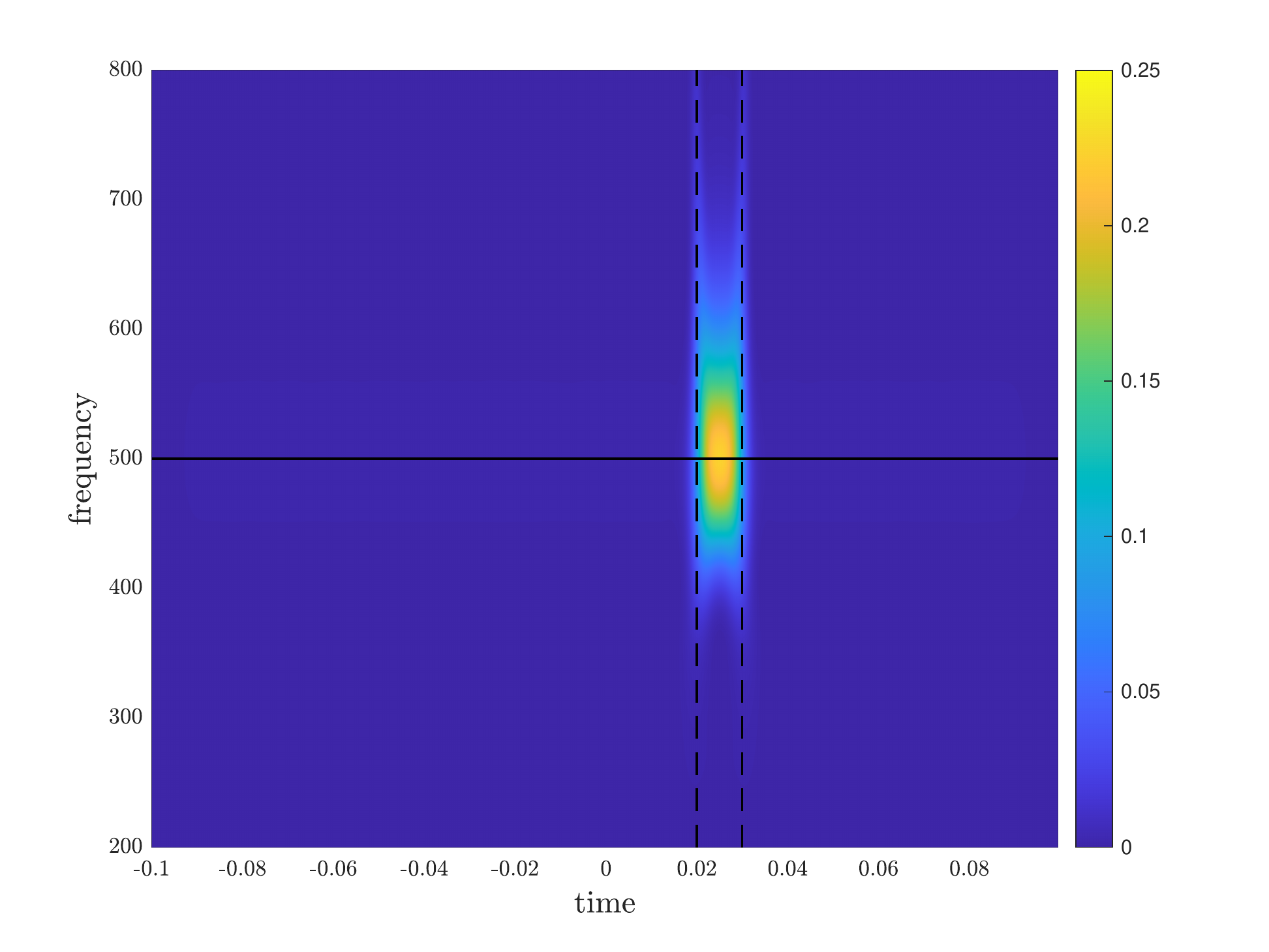} 
  & \includegraphics[width=0.5\linewidth]{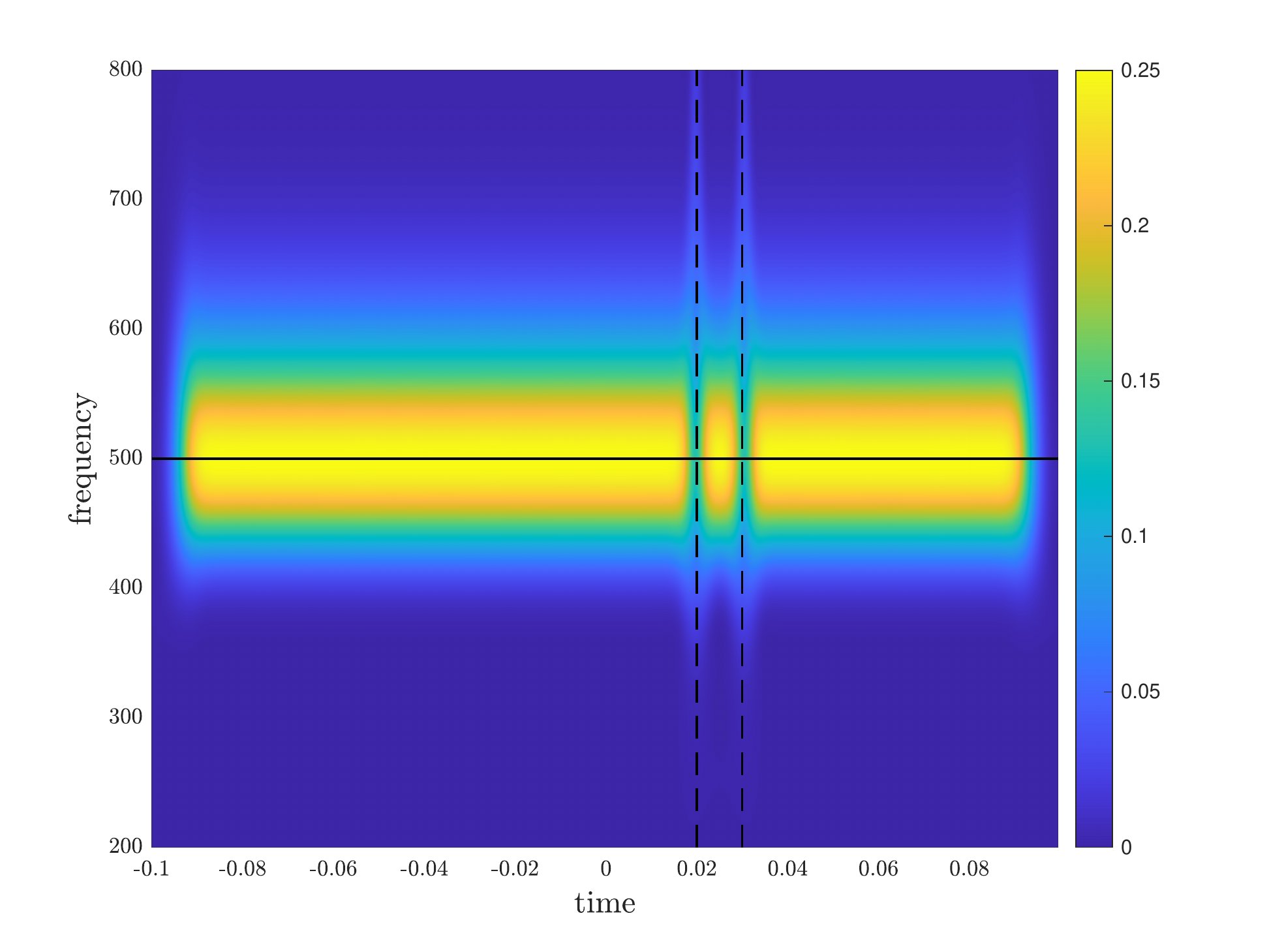} \\
  ITC\textsuperscript{2} & $\mathrm{avgAMP} ^ 2 \times \mathrm{ITC} ^ 2$ as a function of POWavg \\
  \  \\
  \includegraphics[width=0.5\linewidth]{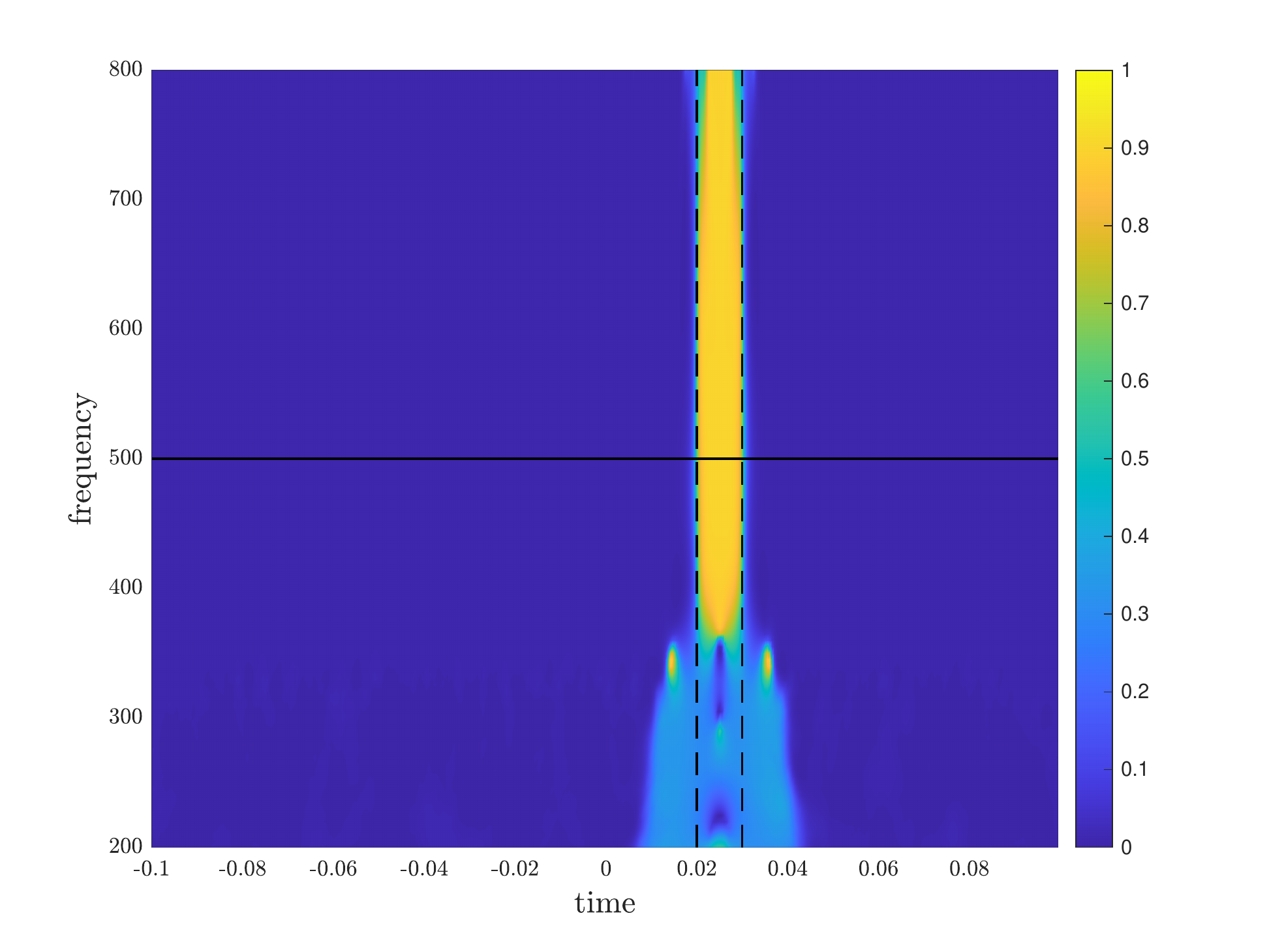}
  & \includegraphics[width=0.5\linewidth]{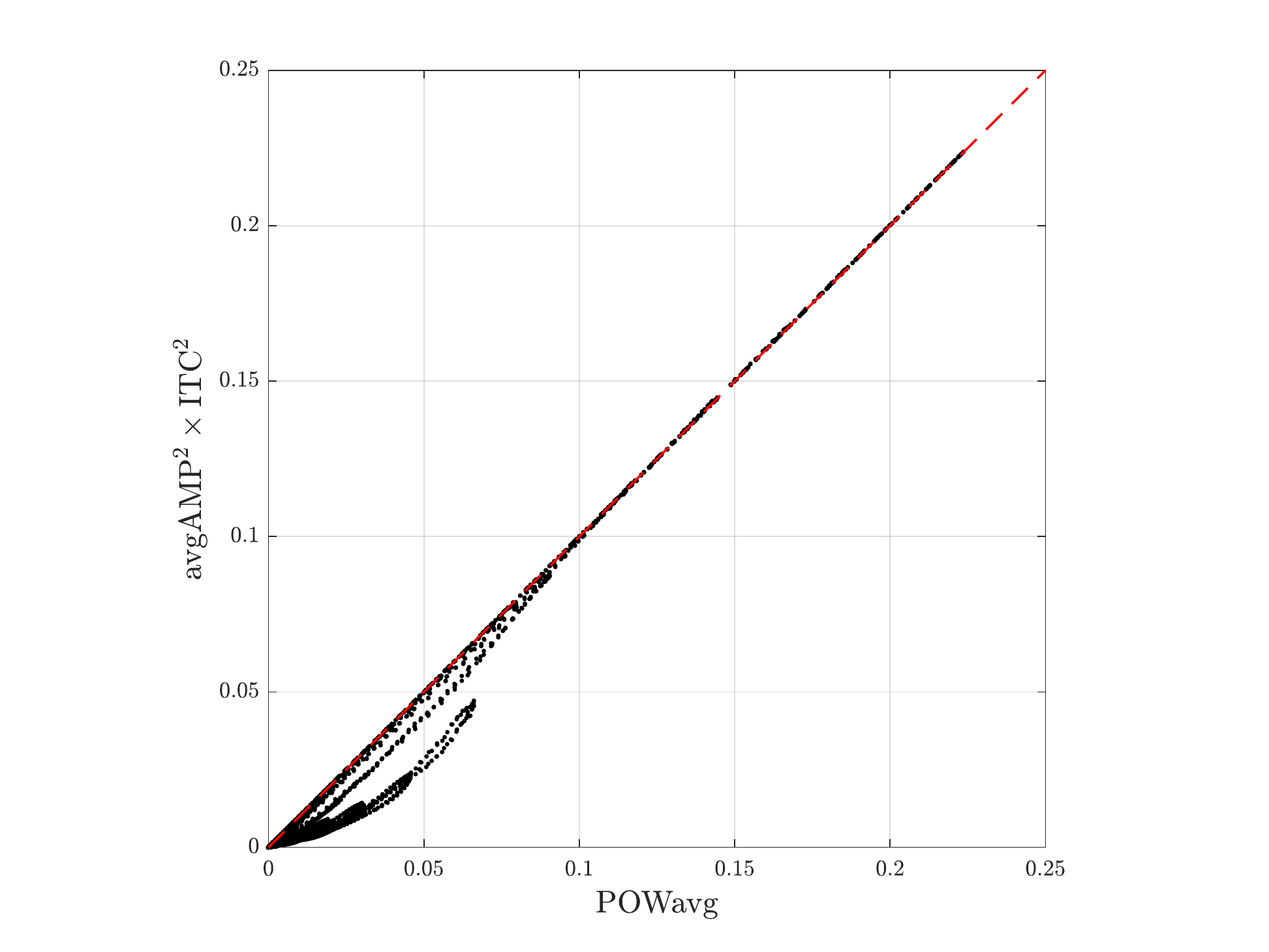} \\
 \end{tabular}
 \caption{\textbf{Simulation study.} Results from \TFT\ with the S-transform for signals with $\kappa^{(i)} = 10$ and $\tau_{\nu} = 0$. Around the expected signal frequency, $f = \nu_0$, the expected value of the power should be close to $\Omega_0 ^ 2 / 4 = 0.25$.}
 \label{fig:simu:res}
\end{figure*}

Results are summarized in Fig.~\ref{fig:simu:res} and in \S1 of \supplmat{2}. avgAMP\textsuperscript{2} evidenced a (mostly) constant signal power around 500~Hz. For an oscillatory signal of the form given by \eqref{eq:osc:m0:def}, the maximum of the power for the S-transform is reached around $f = \nu$, with value close to $\Omega_0 ^ 2 / 4 = 0.25$. The values of avgAMP\textsuperscript{2} were in agreement with this result.
\par
ITC behaved as expected. It was able to detect the increase in phase synchrony between 20 and 30~ms, with very low values (close to 0) outside the 20--30~ms window, and large values (close to 1) within this range. As predicted by our calculations, high values in the 20--30~ms window were not limited to frequencies close to 500~Hz.
\par
POWavg was found to well fit the values predicted by $\mathrm{avgAMP}^2 \times \mathrm{ITC}^2$. As a consequence, it shared the localization properties of both avgAMP and ITC. Like avgAMP, it was larger for frequencies close to 500~Hz; and, like ITC, it was larger in the 20--30~ms window. Both features put together made it so that it was able to precisely localize both the frequency of the signal and the time of increased phase synchronization.
\par
Looking more specifically at the error made by the relationship of \eqref{eq:rel:approx}, we observed that it had a maximum value of about 10\% of the maximum value of POWavg. Its distribution in terms of time and frequency was very specific, as it was lower than 0.005 (corresponding to 2\% of the maximum value of POWavg) for most values of $t$ and $f$, with a relative exception around 20 and 30~ms. These two times corresponded to discontinuities in all measures; see below. Also, largest error values were observed for small values of POWavg: for values of POWavg larger than 0.1 (corresponding to at least 40\% of the maximum), the error was smaller than 0.0025 (corresponding to 1\% of the maximum value of POWavg). The absolute value of the covariance between $e ^ { i \phase{x} }$ and $| \TF{x} |$ was found to follow a distribution in terms of $t$ and $f$ that was similar to that of the error, with a maximal value around 0.08.
\par
These results bring evidence in favor of the validity of the approximations made in the theoretical calculations: (i) approximating measures by their expectations; (ii) using an approximation of the S-transform, \eqref{eq:osc:gen:approx:TFa}; (iii) using asymptotic approximations in $O ( 1/ N )$.

We also noticed a discontinuity in avgAMP\textsuperscript{2} around 20~ms and 30~ms, as well as artifacts of ITC\textsuperscript{2} at lower frequencies. These are probably artifacts of the \TFT\ are due to the sudden change in phase at 20~ms (from $\phi_n^{(o)}$ to $\phi_n^{(i)}$) and at 30~ms (back from $\phi_n^{(i)}$ to $\phi_n^{(o)}$) and limited amount of noise in the data, i.e, artifacts due to the simple model we used.  In particular, the artifacts observed for ITC\textsuperscript{2} were likely a consequence of spectral leakage and were reduced by increasing the standard deviation of the noise.

\graphicspath{{FiguresRevSimus500/}}

\section{Application to experimental data} \label{s:ated}

We rely on \gcitet{Valencia-2006}, who previously investigated the connection between the behaviors of POWavg, avgAMP, and ITC in the context of high-frequency oscillations (HFOs) in early cortical somatosensory evoked potentials.

\subsection{Data}

Somatosensory evoked potentials following median nerve stimulations were recorded in a healthy subject. Brain responses were acquired using multichannel EEG with a sampling frequency of  3~kHz. Electrical median nerve stimulation of 1~ms duration was applied to median nerve at the wrist level to elicit a burst of high-frequency oscillations (HFOs) in the 400--800 frequency range superimposed onto the cortical N20 potential \gcite{Curio-1994b, Curio-1997, Ozaki-1998}. The stimulus was applied 300 times, with a 500-ms inter-trial interval. Following previous recommendations \gcite{Ozaki-1998}, we studied the fronto-central channels (CP3--Fz). Data acquisition was performed at the Center for Neuroimaging Research (CENIR) of the Brain and Spine Institute (ICM, Paris, France). The experimental protocol was approved by the CNRS Ethics Committee and by the national ethical authorities (CPP {\^I}le-de-France, Paris 6 -- Pitié-Salpêtrière and ANSM).

\subsection{Analysis}

The data was analyzed in the window 10--100~ms after stimulation. We started at 10~ms after stimulus to avoid consequences of the artifacts generated by the stimulation. For the \TFT, we used the S-transform.  Akin to what was done for the simulation study, we computed the three measures POWavg, avgAMP\textsuperscript{2}, and ITC\textsuperscript{2}; the error $\left| \mathrm{POWavg} - \mathrm{avgAMP} ^ 2 \times \mathrm{ITC} ^ 2 \right|$; and the absolute value of the covariance between $e ^ { i \phase{x} }$ and $| \TF{x} |$.

\subsection{Results}

\begin{figure*}[!htbp]
  \centering
  \begin{tabular}{cc}
  POWavg & avgAMP\textsuperscript{2} \\
  \includegraphics[width=0.5\linewidth]{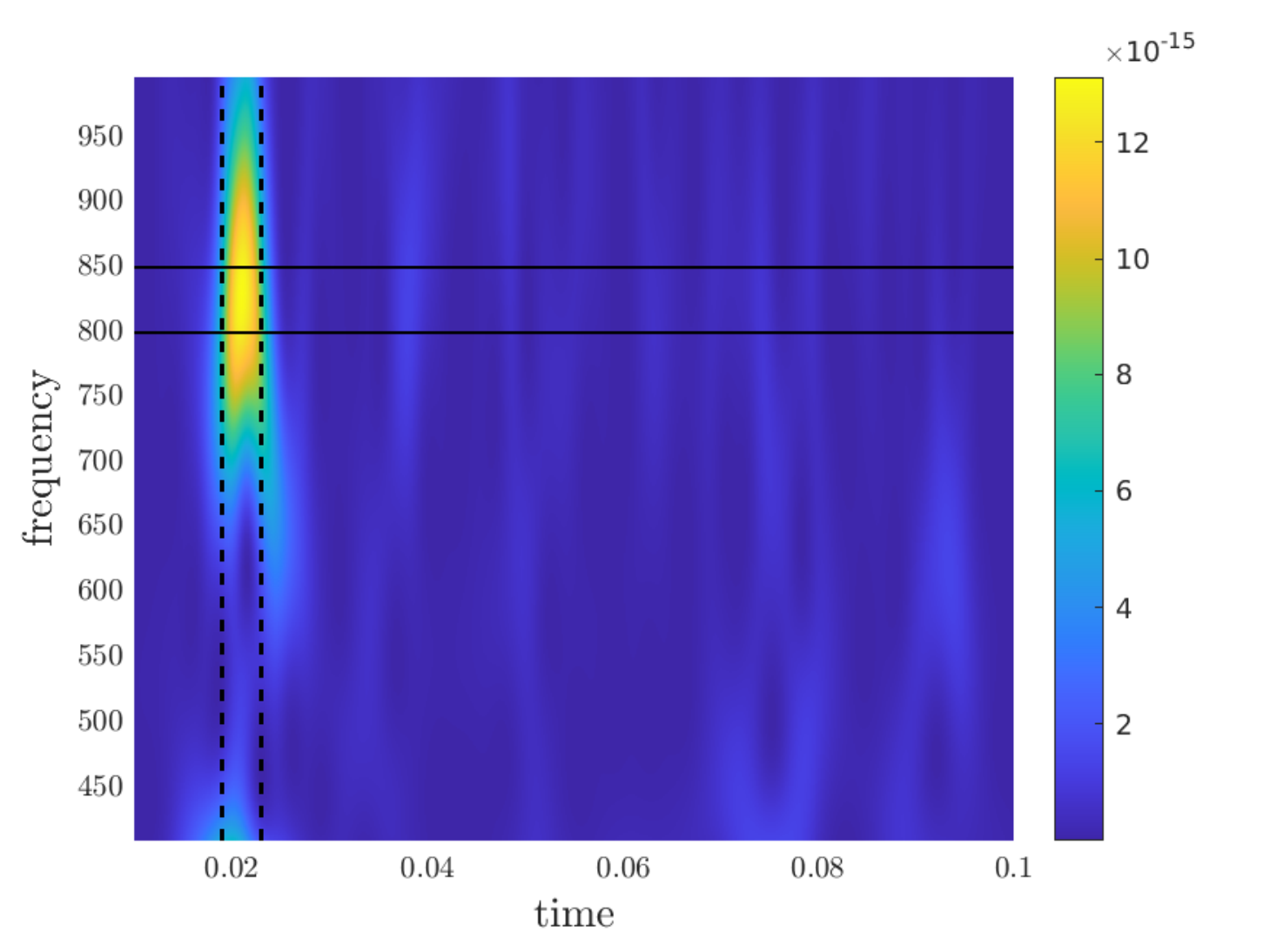} 
  & \includegraphics[width=0.5\linewidth]{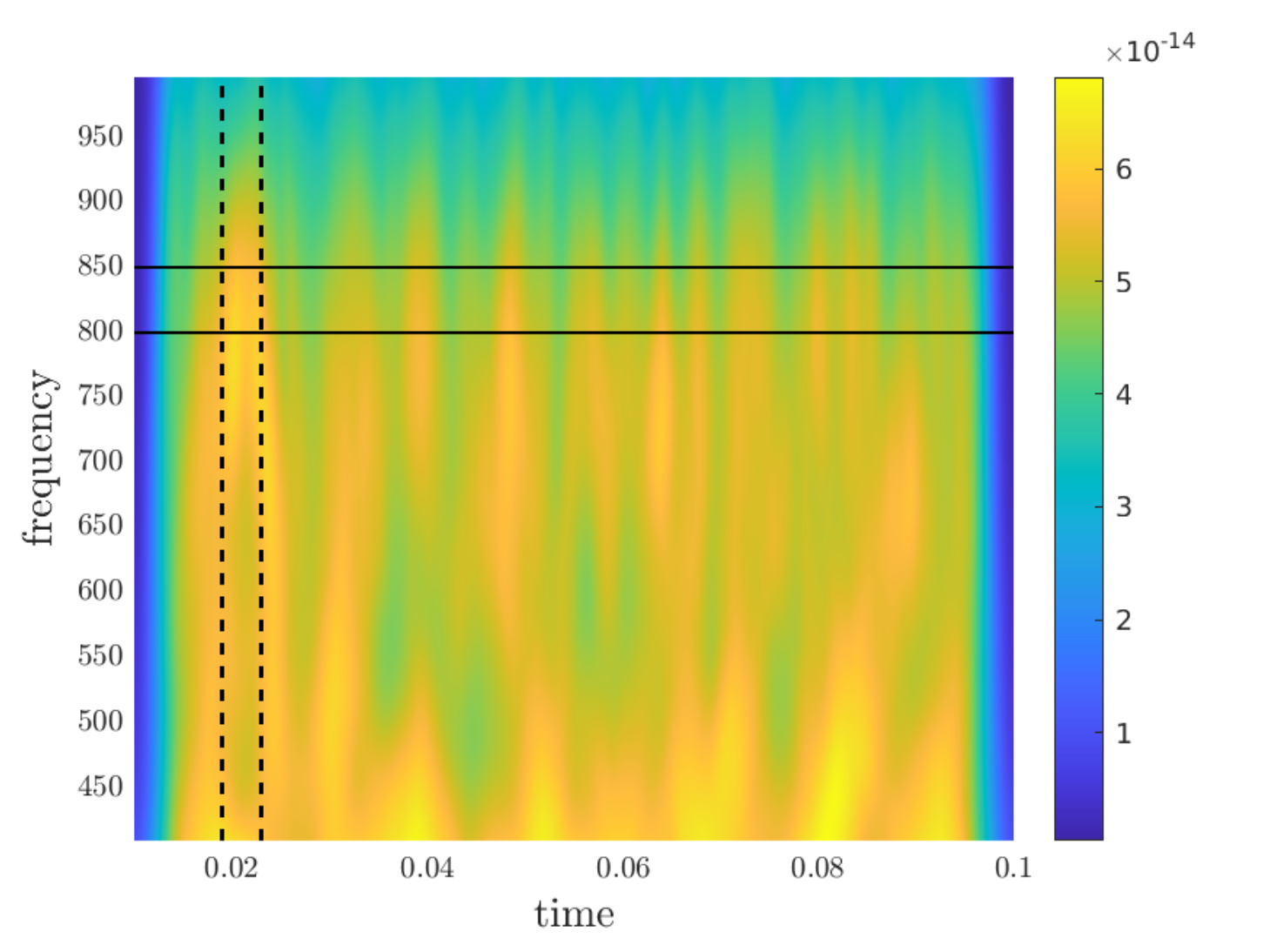} \\
  \ \\
  ITC\textsuperscript{2} & $\mathrm{avgAMP} ^ 2 \times \mathrm{ITC} ^ 2$ as a function of POWavg \\
  \includegraphics[width=0.5\linewidth]{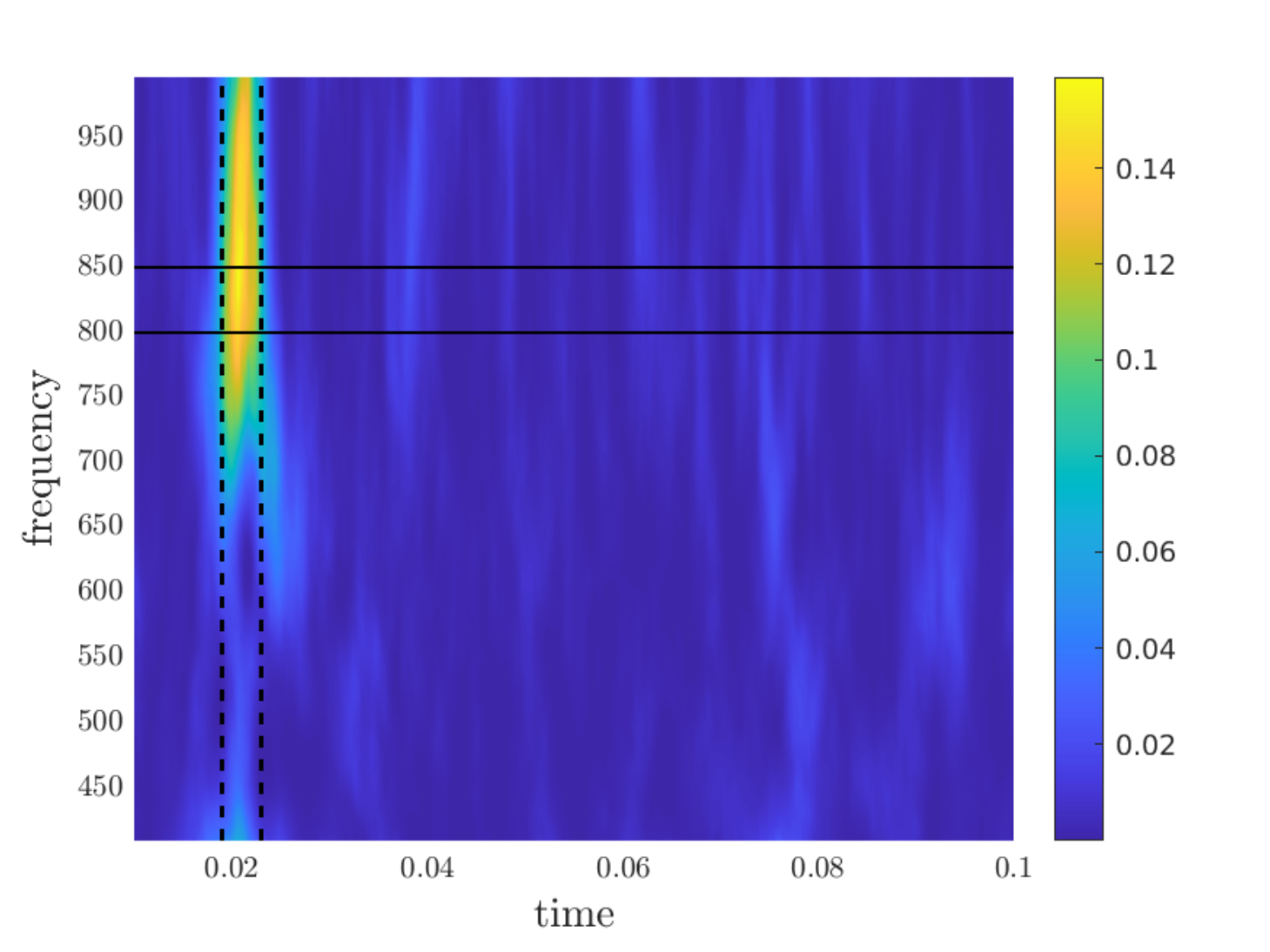} 
  & \includegraphics[width=0.5\linewidth]{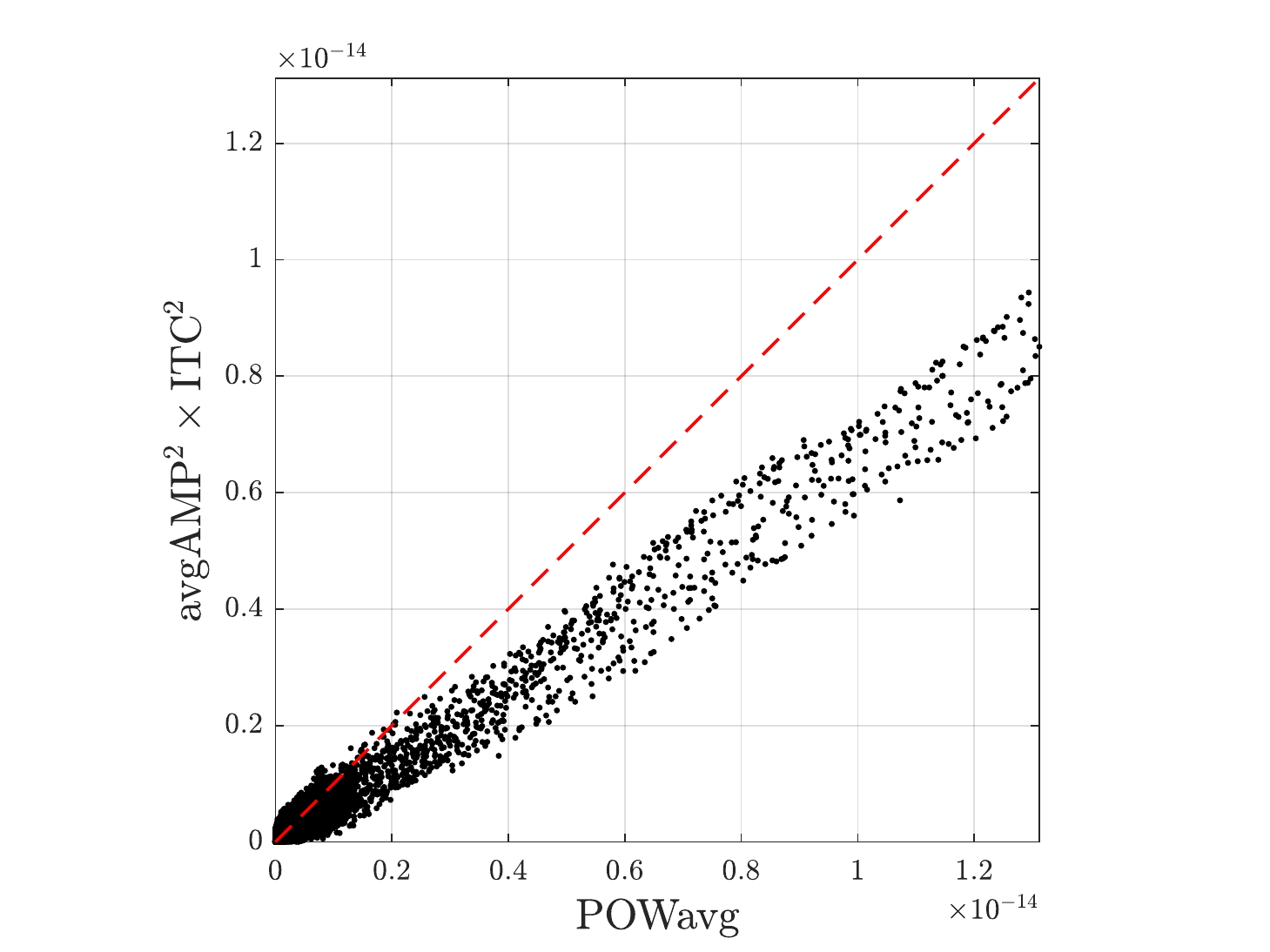} \\
 \end{tabular}
 \caption{\textbf{Real data.} Results from \TFT\ with the S-transform: POWavg, avgAMP\textsuperscript{2}, ITC\textsuperscript{2}, $\left| \mathrm{POWavg} - \mathrm{avgAMP} ^ 2 \times \mathrm{ITC} ^ 2 \right|$, and relation between POWavg and  $\mathrm{avgAMP} ^ 2 \times \mathrm{ITC} ^ 2$. Horizontal solid lines were added at 800 and 850~Hz, and vertical dashed lines were added at 19 and 23~ms.}
 \label{fig:donnees:res}
\end{figure*}

Results are summarized in Fig.~\ref{fig:donnees:res} and in \S2 of \supplmat{2}. avgAMP\textsuperscript{2} did not exhibit a clear frequency band with larger values. POWavg and ITC\textsuperscript{2} both exhibited an increase around 20~ms in the range of 800--850~Hz. We found evidence for a linear relationship between POWavg and $\mathrm{avgAMP}^2 \times \mathrm{ITC}^2$, even though the proportionality factor was found to be rather in the range 0.6--0.7 than close the predicted value of 1. The error made by the approximation of \eqref{eq:rel:approx} was observed to be roughly proportional to POWavg, corresponding to relative error of the order of 30\%, consequently exhibiting larger values where POWavg was larger. The covariance between $e ^ { i \phase{x} }$ and $\left| \TF{x} \right|$ was also observed to be larger in the same time-frequency region.

\graphicspath{{FiguresRevAnalyse/}}

\section{Discussion} \label{s:disc}

In neuroscience, time frequency analysis is a key method for the investigation of brain rhythms from brain recordings, in particular when considering event-related protocols. In this framework, three measures have been proposed, which we defined in \eqref{eq:avgAMP:def}--\eqref{eq:POWavg:def} and coined avgAMP, ITC, and POWavg. While these three measures are sensitive to different features of brain responses, similarities between ITC and POWavg have been observed. In the present manuscript, we further investigated the connection between ITC and POWavg using theoretical calculations and a simulation study. With the theoretical calculations, we derived several variants of a relationship that roughly read POWavg $=$ avgAMP\textsuperscript{2} $\times$ ITC\textsuperscript{2} and whose exact forms depended on the specifics of the underlying models and assumptions. The simulation study nicely confirmed the global form of the relationship, thereby bringing evidence in favor of the validity of the approximations made in the theoretical calculations.
\par
The relationship between POWavg, ITC and avgAMP might seem obvious and highly anticipated, as (i) these three measures are computed from the same brain signals, and (ii) empirical results have showed strong similarities between POWavg and ITC\textsuperscript{2}, e.g., \gcite{Valencia-2006}. Yet, we argue that the results developed in the present manuscript are novel in at least two ways. First, we started with the observed common similarity of results produced using POWavg and ITC and investigated its theoretical underpinning. We then proved that the relationship between these \emph{two} measures actually also involves the \emph{third} measure, avgAMP. We believe that the involvement of avgAMP was not anticipated. The second novelty is that the relationship takes a very simple form, one of proportionality. We do not think that this result was largely expected either.
\par
We advocate that both results are an unlikely coincidence that is the very consequence of the choice of POWavg, ITC and avgAMP as measures of interest. For instance, changing ITC to another measure of angular dispersion, e.g., the circular mean deviation \gcite[\S2.3.4]{Mardia-2000}
\begin{equation}
  d_0 ( \tilde{\phi} ) = \frac{ 1 } { N }\sum_{ n = 1 } ^ N \left[ \pi - \left| \pi - \left| \phase{x_n} - \tilde{\phi} \right| \right| \right],
\end{equation}
where $\tilde{\phi}$ is the sample median of the $\phase{x_n}$'s, would render the relationship invalid. Similarly, changing avgAMP (a measure of amplitude) to a measure of power,
\begin{equation} \label{eq:avgPOW:def}
 \mathrm{avgPOW} = \frac{ 1 } { N } \sum_{ n = 1 } ^ N \left| \TF{x_n} \right| ^ 2,
\end{equation}
would make for a more complex relationship, as we have
\begin{equation}
 \esp ( \mathrm{avgPOW} ) = \esp \left[ \left| \TF{x} \right| \right] ^ 2 + \var \left[ \left| \TF{x} \right| \right],
\end{equation}
to be compared with \eqref{eq:avgAMP2:esp}. In the expression of $\esp ( \mathrm{avgPOW} )$, the variance term does not vanish with increasing $N$, while it does for $\esp ( \mathrm{avgAMP} ^ 2 )$---see \eqref{eq:avgAMP2:asympt}. As a consequence, we have $\mathrm{POWavg} \neq \mathrm{avgPOW} \times \mathrm{ITC} ^ 2$. So it is a subtle interplay between the three measures selected that make the existence of the relationship possible.
\par
As a consequence, the observed similarity between the \emph{two} measures POWavg and ITC actually emphasizes the fact that the \emph{three} measures POWavg, ITC and avgAMP are ``naturally'' connected in ways that remain to be fully understood, and provide three views of the same core time-frequency feature of the signal.
\par
With the calculations, we first investigated the particular case of oscillatory signals (Section~\ref{ss:ioom}). Recordings of brain activity are hypothesized to be interpretable in terms of brain rhythms, i.e., of signals of specific frequency ranges. This is a major reason why \TFT\ is applied to this kind of signals. It is also evidence that oscillatory signals are natural candidates for realist models of brain activity.
\par
To perform the \TFT, we applied the S-transform. There are two reason for this choice. First, it is a method that is commonly applied in the analysis of brain recordings. Furthermore, it had the advantage of rendering our calculations tractable. Its main drawbacks are its (relative) computational burden and the fact that it is not analytic, i.e., its Fourier transform has nonzero power in negative frequencies. As a consequence, it splits the power of the signal on positive and negative frequency components. Our calculations with the S-transform were made possible by replacing the true \TFT\ $\TF{x}$ of \eqref{eq:osc:gen:tf} with the approximation $\TFa{x}$ of \eqref{eq:osc:gen:approx:TFa}. We showed that such an approximation was valid over a wide range of frequencies.
\par
Using theoretical calculations, we also tackled the general case (Section~\ref{ss:ge}). There, we showed that the relationship was of the form of \eqref{eq:rel:esp}, provided that the covariance between the amplitude and phase of the \TFT\ was equal to 0. To reach this conclusion, we relied on the assumption that avgAMP, ITC, and POWavg, which are averages computed over many repetitions $N$ of the same experiment (typically, $N$ is equal to one to several hundreds), could be approximated by expectations. This is equivalent to neglecting the variability of the measures, i.e., equivalent to assuming low variance.
\par
We finally used a simulation study (Section~\ref{s:ss}) to generate synthetic HFOs as observed in event-related protocols involving sensory stimulation. The generative model allowed for a certain level of variability that could not be taken into account in the calculations. Note that we did not simulate the main SEP (N20), as this was not necessary: Since the \TFT\ is linear, the \TFT\ of the SEP would only be superimposed to the \TFT\ of th HFOs. Since frequencies do not overlap, the SEP is expected to have very limited influence on our analysis.
\par
The synthetic data were then analyzed using the S-transform. Despite the above-mentioned variability, the global relationship of \eqref{eq:rel:approx} still seemed to hold relatively well. In particular, the results showed that both ITC and POWavg were sensitive to the simulated phase resetting phenomenon.
\par
Unexpectedly, the results also hinted that ITC might not be able to determine the frequency at which the phase resetting occurs, while POWavg seemed to be able to. This is in agreement with our theoretical calculations---compare, on the one hand, \eqref{eq:osc:m1:ITC}, \eqref{eq:osc:m2:ITC2:esp}, and \eqref{eq:osc:m3:ITC2:esp}, which show that these expressions of ITC (or ITC\textsuperscript{2}, or its expectation) are not a function of $f$, and, on the other hand, \eqref{eq:osc:m1:POWavg}, \eqref{eq:osc:m2:POWavg}, and \eqref{eq:osc:m3:POWavg:esp}, which show that POWavg (or its expectation) are. However, to the best of our knowledge, this is contrary to empirical evidence on real data, where ITC is usually localized in frequency, making possible to narrow the range where the induced oscillations take place \gcite{Tallon-Baudry-1996, Valencia-2006}. We aim to find an explanation to this discrepancy.
\par
Besides the S-transform, another common ways to perform \TFT\ is to use the wavelet transform. Wavelets transforms are fast and, when using analytic wavelets, only decompose a signal in the range of positive frequencies. For instance, the expected power at $f = \nu_0$ of our oscillatory signal is 1 instead of 0.25. In our simulation study, we found that the practical difference between S-transform and wavelet transform was very limited; see, e.g., \S3 of \supplmat{2} for analysis of the synthetic data using wavelet transforms (Morse and analytic Morlet wavelets).
\par
Note that the simulation study used a crude model of brain response. As a consequence, we observed on the simulated data some results that are not typical of  time-frequency diagrams of real brain recordings. Some of these atypical results corresponded to theoretical predictions, while others did not. For instance, as mentioned above, the fact that ITC on simulated HFOs was not localized in a limited frequency band around the true oscillation frequency was to be expected from the theoretical calculations but not from the literature. Other features of the simulated time-frequency diagrams did not correspond to what is typically observed on time-frequency diagrams of real brain recordings. This includes a discontinuity in avgAMP\textsuperscript{2} at 20~ms and 30~ms, as well as artifacts of ITC\textsuperscript{2} at lower frequencies. These are probably artifacts of the \TFT\ might be due to the sudden change in phase at 20~ms (from $\phi_n^{(o)}$ to $\phi_n^{(i)}$) and at 30~ms (back from $\phi_n^{(i)}$ to $\phi_n^{(o)}$) and limited amount of noise in the data. As a consequence, the possibility that we had in our simulations to detect phase resetting in avgAMP is merely an artifact resulting from an oversimplified simulation model rather than an empirically confirmed result.
\par
While we focused on HFOs for the simulation study, nothing in the theoretical developments depends on the frequency of the underlying signal. As a consequence, we expect the general relationship to hold for different brain frequency ranges as well. As a confirmation, we ran a new simulation with $\nu_0 = 40$~Hz. As expected, the results were very similar to the ones obtained for the simulated HFOs (for details, see \S4 of \supplmat{2}). This is particularly relevant in the context of auditory evoked steady-state responses (ASSR) which are increasingly being used to probe the ability of auditory circuits to produce synchronous activity at 40~Hz, in response to repetitive external stimulation \gcite{Galambos-1981}. Perturbation in both power or phase-locking has been also observed in various neuropsychiatric disorders \gcite{Kwon-1999, Wilson_TW-2007}. Because the ASSR is composed of complex time-dependent frequency responses \gcite{Tada-2021}, the present study suggests that the combination of different methods (avgAMP, ITC, and POWavg) can be used to avoid a misleading or wrong interpretation of the neuronal modulations, especially as markers of brain dysfunctions in psychiatric disorders.
\par
Also, the present work focused on the case of a signal composed of a single frequency. We expect the presence of multiple frequencies to have a limited impact on the calculation of avgAMP beyond mere additivity, but to strongly interfere with the calculation of ITC, potentially reducing the frequency range over which a phase resetting phenomenon is associated with a larger value of ITC. Further simulation studies could help assess the validity of such an assumption. It would also be interesting to investigate the particular case where there exists dependencies between frequencies, e.g., in the form of cross-frequency coupling.
\par
From the theoretical calculations, we found that the relationship \eqref{eq:rel:approx} between POWavg, avgAMP, and ITC requires that amplitude and phase have 0 covariance. This requirement is slightly more general than independence, as there exist variables that have 0 covariance but are not independent. On the simulated data, we quantified the level of validity of this assumption and found a rather good fit, at the exception of around 20 and 30~ms, where we observed spectral leakage. On the experimental data, we found the presence of a linear relationship between quantities, albeit with a proportionality factor that was lower than the expected value of 1. Also, the values of both $| \mathrm{POWavg} - \mathrm{avgAMP} ^ 2 \times \mathrm{ITC} ^ 2 |$ and $| \cov [ e ^ { i \phase{x} }, | \TF{x} | ] |$  were roughly linearly related to the values of POWavg. We hypothesize that the departure from theoretical expectations might be related to the presence of noise.
\par
Regarding noise, it has to be mentioned that the present work is based on the assumption that there is no noise and that the totality of the signal measured is relevant. This is usually not the case, as brain recordings are subject to noise. The influence of this noise on \TFT\ needs to be further investigated. In particular, we need to investigate whether it could be the source of the ``dampening'' effect on the linear relationship observed in the real data. Note that, by contrast, \gcitet[\S19.3]{Cohen_MX-2014} expects noise to increase ITC. We hope to be able to clarify this point in the near future. In this endeavor, the relationship of \eqref{eq:rel:approx} might prove to be a valuable tool.
\par
We can now conclude regarding the implication of the present development to neuroscientific practice. Time-frequency analyses have been widely used in electrophysiological research for the characterization of oscillatory responses to stimuli in a large range of brain processes, from simple sensory processing to higher order cognition \gcite{Tallon-Baudry-1999, Varela-2001, Makeig-2004}. Multiple studies have reported evidence that evoked oscillations are not exclusively composed of either a stimulus phasic-related component or some stimulus-induced phase resetting, but usually include a combination of both in variable proportions \gcite{Shah_AS-2004, Hanslmayr-2007}. As a consequence, the respective contributions of both types of components have to be clarified before attempting to draw physiological conclusions. Unfortunately, distinguishing between the two phenomena remains quite a challenge, in particular when multiple oscillations are present in neighboring frequency bands \gcite{Yeung_N-2004, van_Diepen-2018}. One potentially fruitful approach for future research could be to develop measures quantifying to what extent changes in ITC and in avgAMP are correlated in the time and frequency domains. The exact bearing of the relationship will become increasingly clearer as real data are analyzed with this result in mind.

\section{Conclusion}

In this manuscript, we showed that avgAMP, ITC, and POWavg were related through the approximate, simple relationship given by \eqref{eq:rel:approx}. This result, based on theoretical calculations and a simulation study, confirms previous empirical evidence and provides a novel perspective to investigate evoked brain rhythms. In particular, looking at ITC and avgAMP as two distinct measures quantifying different aspects of rhythmic brain activity, and taking into consideration their level of correlation may help disentangle contributions from additive evoked activity and phase-reset oscillatory activity. In this context, the presented relationship between POWavg, ITC, and avgAMP may provide a significant refinement to the neuroscientific toolbox for studying evoked oscillations.
\par
As a final note, time-frequency analysis is a method that is commonly used in many fields of scientific research, including fluids mechanics, engineering, other fields of biomedical imaging and medicine, and geophysics \gcite{Addison-2002}. We hope that the present development will find applications in these fields as well.

\appendices


\section*{Acknowledgment}

The authors would like to thank the Center for Neuroimaging Research (CENIR) of the Brain and Spine Institute (ICM, Paris, France) for the acquisition of the experimental data, and Véronique Marchand-Pauvert for providing them with the data.

\ifCLASSOPTIONcaptionsoff
  \newpage
\fi

\end{document}